\documentclass[a4paper,10pt]{article}
\usepackage[latin1]{inputenc}    
\usepackage[T1]{fontenc}

\usepackage[english]{babel}

\usepackage{graphicx} 

\usepackage{amsmath}
\usepackage{amssymb} 
\usepackage{verbatim}
\usepackage{bm}
\title{}
\author{}

\newcommand{\ii}{\text{i}}

\newtheorem{prop}{Proposition}[section]

\numberwithin{equation}{section}

\begin{document}

\title{Global gauge anomalies\\ in coset models of conformal field theory} 
\author{Paul de Fromont\\{\small{95 cours Vitton, 69006 Lyon, France}}\\{}\\
Krzysztof Gaw\c{e}dzki\\
{\small{Laboratoire de Physique, C.N.R.S., ENS de Lyon,
Universit\'e de Lyon,}}\\\small{46 All\'ee d'Italie, 69364 Lyon, France}\\{}\\
Cl\'ement Tauber\\
{\small{Laboratoire de Physique, ENS de Lyon,
Universit\'e de Lyon,}}\\\small{46 All\'ee d'Italie, 69364 Lyon, France}}

\date{}

\maketitle
\vskip 0.1cm

\centerline{\small\bf Abstract}
\vskip 0.15cm

\parbox[t]{14cm}{\small{\noindent We study the occurrence of global gauge 
anomalies in the 
coset models of two-dimensional conformal field theory that are based 
on gauged WZW models. A complete classification of the non-anomalous theories 
for a wide family of gauged rigid adjoint or twisted-adjoint symmetries
of WZW models is achieved with the help of Dynkin's classification of
Lie subalgebras of simple Lie algebras.}}
\vskip 1.3cm

\section{Introduction}
\label{sec:1}

Bosonic sigma models with the metric action functional 
possess rigid symmetries induced by isometries
of their target space. Such rigid symmetries may be gauged
by the minimal coupling to the gauge fields of the isometry
group. The gauged action is then invariant under arbitrary 
local gauge transformations. The minimal coupling does not work, 
however, for the topological Wess-Zumino term in the action 
functional of the sigma model, if such is present. In particular,
it was shown in \cite{JJMO,HS} for the two-dimensional sigma model 
with the Wess-Zumino term corresponding to a closed $3$-form $H$ 
on the target space that the gauging of rigid symmetries 
requires satisfying certain conditions. Such conditions assure
the absence of local gauge anomalies and guarantee the existence
of a gauging procedure that results in an action functional invariant 
under infinitesimal local gauge transformations.  
The infinitesimal gauge invariance of the gauged action implies 
its invariance under all ``small'' local gauge transformations, 
i.e. the ones that are homotopic to unity. As was observed in
\cite{GSW}, it is possible, however, that the gauged action
exhibits global gauge anomalies that lead to its non-invariance 
under some ``large'' local gauge transformations non-homotopic 
to unity. The phenomenon was analyzed in detail for sigma models 
on closed worldsheets in \cite{GSW} and on worldsheets
with boundaries and defects in \cite{GSW1}. In the case of Wess-Zumino-Witten 
(WZW) models of conformal field theory with Lie group $G=\tilde G/Z$ as the
target, where $\tilde G$ is the universal covering group of $G$ and $Z$ 
is a subgroup of the center $\tilde Z$ of $\tilde G$, 
with the Wess-Zumino term corresponding to the bi-invariant 
closed 3-form $H_k=\frac{k}{12\pi}\,{\rm tr}\hspace{0.03cm}(g^{-1}dg)^3$, 
the local gauge anomalies are absent for a restricted class
of rigid symmetries. These include the symmetries induced 
by the adjoint action $g\mapsto hgh^{-1}$ on $G$ for 
$h\in\tilde G/\tilde Z$, or by its twisted versions 
$g\mapsto hg\hspace{0.03cm}\omega(h)^{-1}$, for 
$h\in\tilde G/Z^\omega$, where $\omega$ is an automorphism
of $\tilde G$ and $Z^\omega=\{z\in\tilde Z\,|\,z\hspace{0.02cm}
\omega(z)^{-1}\in Z\}$
is the subgroup of elements in $\tilde Z$ that acts trivially. 
In these cases, the global gauge anomalies may occur for the target 
groups $G$ that are not simply connected (corresponding to the so called 
non-diagonal WZW models). They are detected by a cohomology class 
$\varphi\in H^2(\tilde G/Z^\omega\times G,U(1))$ that can be easily 
computed. Class $\varphi$ is invariant under the 
action of $\gamma\in\tilde G/Z^\omega$ on $\tilde G/\tilde Z^\omega
\times G$ given by 
\begin{equation}
(h,g)\ \mapsto\ (\gamma h\gamma^{-1},\gamma g\hspace{0.04cm}
\omega(\gamma)^{-1})\,.
\label{ac}
\end{equation} 
The simplest case when the anomaly class is nontrivial corresponds 
to $G=SU(3)/\mathbb{Z}_3$ at level $k=1$ or to $G=SU(4)/\mathbb{Z}_4$ 
at level $k=2$, both with $\omega={I\hspace{-0.04cm}d}$. Some other 
cases with global gauge anomalies for $\omega={I\hspace{-0.04cm}d}$ 
were cited in \cite{GSW}. In Sec.\,\ref{sec:3} of the present paper, 
we obtain the full list of connected compact simple target groups $G$ 
for which the WZW model with the gauged (twisted) adjoint action of 
$\tilde G/\tilde Z^\omega$ exhibits global gauge anomalies. In the twisted 
case, we consider only outer automorphisms $\omega$ since for 
inner automorphisms the twisted adjoint action may be 
reduced to the untwisted one by conjugating it with a right 
translation on $G$ which is a rigid symmetry of the WZW theory.
The classes of outer automorphisms of $\tilde G$ modulo inner automorphisms
are generated by automorphisms of the Lie algebra $\mathfrak{g}$ that
preserve the set of simple roots inducing a symmetry of the Dynkin 
diagram of $\mathfrak{g}$.
Global gauge anomalies occur only for (non-simply connected) 
groups $G$ with Lie algebras $\mathfrak{g}=A_r,\,D_r,\,\mathfrak{e}_6$ 
in the Cartan classification of simple Lie algebras\footnote{We consider 
the compact real forms $\mathfrak{g}$ of complex simple Lie algebras that 
are in one-to-one correspondence with their complexifications 
$\mathfrak{g}^{\mathbb C}$.}.  
\vskip 0.1cm

Gauged WZW models serve to construct coset $G/H$ models \cite{GKO,Godd} 
of the two-dimensional conformal field theory \cite{BRS,GK0,GK,KPSY}. 
In such models, 
one restricts the gauging to the (possibly twisted) adjoint action on 
the target group of the subgroup 
$\Gamma=\tilde H/(Z^\omega\cap\tilde H)\subset
\tilde G/Z^\omega$, where $\tilde H$ a closed 
connected subgroup of $\tilde G$ (simply-connected or not). Global gauge 
anomalies are now detected by the pullback cohomology class 
in $H^2(\Gamma\times G,U(1))$. 
Secs.\,\ref{sec:4} and \ref{sec:5} are devoted to finding out when the latter 
is nontrivial for groups $\,G\,$ as before and for a wide class of 
subgroups $\tilde H\subset\tilde G$ (the nontriviality of the pullback 
class depends only on the subgroup $\tilde H$ modulo conjugation 
by elements of $\tilde G$ and it may occur only if the original anomaly class 
$\varphi$ is nontrivial, hence for Lie algebras $\mathfrak{g}$ 
enumerated above). Closed connected subgroups $\tilde H\subset\tilde G$
are in one-to-one correspondence to Lie subalgebras $\mathfrak{h}\subset
\mathfrak{g}$. We obtain the complete list of cases with global 
gauge anomalies for subgroups $\tilde H$ with the Lie algebra $\mathfrak{h}$ 
which is a semisimple regular subalgebra of $\mathfrak{g}$ (i.e. such that 
the roots of $\mathfrak{h}$ form a  subset of roots of $\mathfrak{g}$). 
The complete classification (modulo conjugation) of regular subalgebras 
of simple Lie algebras was obtained in the classical work \cite{Dynkin}
of Dynkin. The complete classification of all semisimple subalgebras 
of simple Lie algebras is not known explicitly, except for low ranks 
and may be complicated. We give the complete list of non-regular 
semisimple subalgebras $\mathfrak{h}$ of $\mathfrak{g}=\mathfrak{e}_6$
corresponding to subgroups $\tilde H\subset\tilde G$ that lead to global 
gauge anomalies. For $\mathfrak{g}=A_r$ 
and $\mathfrak{g}=D_r$, we limit ourselves to few examples of anomalous 
subgroups $\tilde H\subset\tilde G$ for which $\mathfrak{h}$ is a non-regular 
semisimple subalgebra of $\mathfrak{g}$.
\vskip 0.1cm

As discussed in \cite{GSW} for the untwisted case, the presence of global 
gauge anomalies of the type studied here renders the $G/H$ coset models 
inconsistent on the quantum level (barring accidental degeneracies 
of the affine characters). Hence the importance of the classification 
of the anomalous cases.

\section{No-anomaly condition}
\label{sec:2}

The WZ contribution to the action of the WZW model corresponding 
to the closed $3$-form $H_k$ on a connected compact simple Lie group $G=\tilde G/Z$ with $Z\subset\tilde Z$ may be defined (modulo $2\pi$) whenever 
the periods of $H_k$ (i.e.
its integrals over closed $3$-cycles) belong to $2\pi\mathbb Z$. 
For the standard 
normalization of the invariant negative-definite quadratic 
form ${\rm tr}$ on the Lie algebra 
$\mathfrak{g}$ in which long roots (viewed as elements of 
$\ii\mathfrak t_{\mathfrak g}$, where $\mathfrak t_{\mathfrak g}$ is the
Cartan subalgebra of $\mathfrak g$) have length squared $2$, this happens 
for levels $k\in K_G\subset\mathbb Z$. If $G=\tilde G$ then
$K_G=\mathbb Z$ whereas $K_G$ may be a proper subset 
of $\mathbb Z$ if $G=\tilde G/Z$ with $Z$ nontrivial (i.e. $\not=\{1\}$). 
Sets $K_G$ of admissible levels are explicitly known 
\cite{FGK,Gawedzki}. Besides, for $G=SO(2r)/{\mathbb Z_2}$ with $r$ even 
(where $K_G=\mathbb{Z}$ when $4|r$ and $K_G
=2\mathbb{Z}$ if $4\hspace{-0.2cm}\not{\hspace{-0.05cm}|}r$), there are 
two different consistent choices of the WZ term of the action. 
The details of the construction of the WZ contribution 
$\exp\big[iS^{W\hspace{-0.05cm}Z}_\Sigma(g)\big]$ to the Feynman 
amplitude of the sigma-model field $g:\Sigma\to G$ defined on a closed 
oriented worldsheet $\Sigma$, discussed e.g. in \cite{Gtop,GR}, will not 
interest us here beyond the fact that the result is invariant under 
the composition of fields $g$ with the left or right action of (fixed) 
elements of group $G$. The action functional with the (twisted) adjoint 
symmetry of the WZW model gauged is a functional of field $g$ and of 
gauge-field $A$, a $\mathfrak{g}$-valued $1$-form on $\Sigma$. 
It has the form
\begin{equation}
S^{W\hspace{-0.05cm}Z}_\Sigma(g,A)\,=\,S^{W\hspace{-0.05cm}Z}_\Sigma(g)\,
+\,\frac{_k}{^{4\pi}}\int{\rm tr}\,\big((g^{-1}dg)\hspace{0.02cm}\omega(A)
+(dg)g^{-1}\hspace{-0.05cm}A+g^{-1}\hspace{-0.05cm}
Ag\hspace{0.05cm}\omega(A)\big)
\label{gauged_action}
\end{equation}
(for the untwisted case, $\omega=I\hspace{-0.03cm}d$).
The local gauge transformations $h:\Sigma\to\tilde G/Z^\omega$ act 
on the sigma model and gauge fields by
\begin{equation}
{}^h\hspace{-0.05cm}g\,=\,hg\hspace{0.04cm}\omega(h)^{-1}\,,
\qquad{}^h\hspace{-0.06cm}A\,
=\,hA\hspace{0.01cm}h^{-1}+hdh^{-1}\,.
\end{equation}
Note that $Z^\omega=\tilde Z$ for $\omega=I\hspace{-0.03cm}d$.
It is easy to show that the invariance of the gauged Feynman amplitudes 
under such transformations: 
\begin{equation}
\exp\big[iS^{W\hspace{-0.05cm}Z}_\Sigma({}^h\hspace{-0.05cm}g,
{}^h\hspace{-0.06cm}A)
\big]\,=\,\exp\big[iS^{W\hspace{-0.05cm}Z}_\Sigma(g,A)\big]
\label{gaugeinv}
\end{equation}
is equivalent to the identity
\begin{eqnarray}
\frac{\exp\big[iS^{W\hspace{-0.05cm}Z}_\Sigma({}^h\hspace{-0.05cm}g)\big]}
{\exp\big[iS^{W\hspace{-0.05cm}Z}_\Sigma(g)\,
+\,\frac{_{\ii k}}{^{4\pi}}\int_\Sigma
{\rm tr}\,\big(g^{-1}dg\hspace{0.05cm}\omega(h^{-1}dh)+(dg)g^{-1}h^{-1}dh
+g^{-1}(h^{-1}dh)g\hspace{0.05cm}\omega(h^{-1}dh)\big)\big]}\,
=\,1\,,\  
\label{eval}
\end{eqnarray}
see Appendix \ref{app:1}.
The ratio on the left hand side belongs always to $U(1)$. It coincides 
with the evaluation of the anomaly class $\varphi\in 
H^2(\tilde G/Z^\omega\times G,U(1))$ on the 2-cocycle that is the image of the 
fundamental class of $\Sigma$ under the map $(h,g):\Sigma\to\tilde G/
Z^\omega\times G$.

A simple analysis \cite{GSW}
of the structure of cohomology group $H^2(\tilde G/Z^\omega\times G,U(1))$ 
based on the K\"unneth Theorem shows that class $\varphi$ is trivial 
if and only if
identity (\ref{eval}) holds for $\Sigma=S^1\times S^1$ and 
\begin{equation}
h({\rm e}^{\ii\sigma_1},{\rm e}^{\ii\sigma_2})={\rm e}^{\ii\sigma_1\tilde M}\,,
\qquad g({\rm e}^{\ii\sigma_1},
{\rm e}^{\ii\sigma_2})={\rm e}^{\ii\sigma_2M}
\label{hg}
\end{equation}
where $\tilde M,M\in\ii\mathfrak t_{\mathfrak g}$ and are such that, 
in terms of the exponential map with values in $\tilde G$,
\begin{equation} 
\tilde z\equiv e^{2\ii\pi\tilde M}\in Z^\omega\quad\ {\rm and}
\quad\ z\equiv e^{2\ii\pi M}\in Z\,.
\label{tzz}
\end{equation}
Both $\tilde M$ and $M$ have to belong to the coweight lattice 
$P^\vee(\mathfrak{g})\subset\ii\mathfrak{t}_{\mathfrak g}$ dual to the
weight lattice of $\mathfrak{g}$ and composed of 
$M\in\ii\mathfrak{t}_{\mathfrak g}$ s.t. $\exp[2\ii\pi M]\in\tilde Z$.
For $(h,g)$ given by Eqs.\,(\ref{hg}), the left hand side of Eq.\,(\ref{eval})
is easily computable giving rise to the identity
\begin{equation}
c_{\tilde z\omega(\tilde z)^{-1},z}\,\exp\big[-2\ii\pi k\,
{\rm tr}(M\omega(\tilde M))\big]\,=\,1
\label{idtytw}
\end{equation}
which holds for all $\tilde M,M\in P^\vee(\mathfrak{g})$ 
as above if and only if there
are no global gauge anomalies for the WZW model with gauged (twisted) adjoint 
action of $\tilde G/\tilde Z^\omega$ on the target group $G$. 
In Eq.\,(\ref{idtytw}),
\begin{equation}
Z^2\ni(z,z')\,\mapsto\,c_{z, z'}\in U(1) 
\label{bihom}
\end{equation}
is a $k$-dependent bihomomorphism in $Hom(Z\otimes Z,U(1))$
whose explicit form may be extracted from Appendix 2 of \cite{FGK}.
For cyclic $Z\equiv\mathbb Z_p$ generated by $z_0=e^{2\ii\pi\theta}$
for $\theta\in P^\vee(\mathfrak{g})$,
\begin{equation}
c_{z_0^m,z_0^n}=\exp[-\ii\pi k\hspace{0.02cm}mn\,{\rm tr}(\theta^2)].
\label{cc}
\end{equation}
For the only case with non-cyclic $Z$, we shall explicit $c_{z,z'}$
in Sec.\,\ref{sec:D_reven_tw}. In the untwisted case with 
$\omega=I\hspace{-0.03cm}d$, 
condition (\ref{idtytw}) reduces to the requirement that
\begin{equation}
\exp\big[-2\ii\pi k\,{\rm tr}(M\tilde M)\big]\,=\,1\,.
\label{idty}
\end{equation}
If we gauge only the adjoint action of $\tilde H/(\tilde Z^\omega
\cap\tilde H)$ 
then there are no global gauge anomalies if and only if identity (\ref{idtytw}) 
holds under the additional restriction that, as an element of $\tilde G$, 
$\,\exp[2\ii\pi\tilde M]\in\tilde H$. 
\vskip 0.1cm

It is enough to check the above conditions for $\tilde M,M$ in different
classes modulo the coroot lattice $Q^\vee(\mathfrak{g})$ (composed
of $\tilde M\in\ii\mathfrak{t}_{\mathfrak g}$ s.t. $\exp[2\ii\pi\tilde M]=1$
in $\tilde G$) since ${\rm tr}\,\tilde M M\in{\mathbb Z}$ if 
$\tilde M\in P^\vee(\mathfrak{g})$ and $M\in Q^\vee(\mathfrak{g})$ 
or {\it vice versa}. In particular, if $Z=\{1\}$, i.e. if $G$ is simply
connected, then conditions (\ref{idtytw}) and (\ref{idty}) are 
always satisfied so that there are no global gauge anomalies in that case. 
In the sequel, we shall describe for each Lie algebra $\mathfrak{g}$ 
the center $\tilde Z$ 
of the corresponding simply connected group $\tilde G$ in terms 
of coweights of $\mathfrak{g}$. Then choosing a Lie 
subalgebra $\mathfrak{h}\subset\mathfrak{g}$, we shall restrict elements 
$\tilde{M}$ by requiring that $e^{2\ii \pi \tilde{M}}\in \tilde{H}$. 
Note that $e^{2\ii\pi\tilde M}\in\tilde H\,$ 
if and only if $\,e^{2\ii\pi\tilde M}\in g\tilde H g^{-1}\,$ 
for $\,g\in\tilde G$ and $\,e^{2\ii\pi\tilde M}\in\tilde Z$. 
Hence the no-anomaly conditions coincide for conjugate subgroups 
$\tilde H\subset\tilde G$. Thus it is 
enough to consider one Lie subalgebra $\,\mathfrak h\subset\mathfrak g\,$ 
in each class of subalgebras related by inner automorphisms of $\mathfrak g$. 
We may also require that the Cartan subalgebra $\mathfrak t_{\mathfrak h}$
of $\mathfrak h$ be contained in the Cartan subalgebra 
$\mathfrak t_{\mathfrak g}$ of $\mathfrak g$. Then 
$e^{2\ii \pi \tilde{M}}\in \tilde{H}$ if and only if there is 
$q^\vee\in Q^\vee(\mathfrak g)$ such that $\tilde M+q^\vee\in\ii\mathfrak 
t_{\mathfrak h}$. This is the condition that we shall impose 
on $\tilde M$. 
\vskip 0.1cm

The no-anomaly conditions for Lie subalgebras $\mathfrak h\subset\mathfrak g$
related by outer automorphisms $\omega'$ of $\mathfrak g$ are also related.
Indeed, it is easy to see that the expression on the right hand side
of Eq.\,(\ref{eval}) for gauge transformation $h$ and fields $g$ coincides 
with the similar expression for gauge transformation $\omega'(h)$ and 
field $\omega'(g)$ if in the latter case subgroup $Z\subset\tilde Z$ is 
replaced by $\omega'(Z)$ and the twist $\omega$ by 
$\omega'\omega\hspace{0.04cm}\omega'^{-1}$. The only exception is 
the case of $G=SO(2r)/\mathbb{Z}_2$ for even $r$ and odd $k$ where 
one may also have to interchange the two different consistent choices 
of the theory, see Sec.\,\ref{sec:D_reven_tw}.
\vskip 0.2cm

{\bf Summarizing:} the necessary and sufficient condition for 
the absence of global gauge anomalies requires that Eq.\,(\ref{idtytw})
holds for all $\tilde M,M\in P^\vee(\mathfrak{g})$ such that
\begin{equation} 
\tilde z\equiv e^{2\ii\pi\tilde M}\in Z^\omega\cap\tilde H\quad\ {\rm and}
\quad\ z\equiv e^{2\ii\pi M}\in Z\,.
\label{tzzH}
\end{equation}
In the untwisted case, this reduces to the condition\footnote{In the conformal
filed theory terminology \cite{SchellYank}, condition
(\ref{coset_anomaly}) means that the monodromy charge $Q_J(\tilde J)$ for the 
simple currents $\tilde J$ and $J$ corresponding to the central elements 
$\tilde z$ and $z$ has to vanish modulo 1.}
\begin{equation}\label{coset_anomaly}k~\text{tr} 
(M \tilde{M}) \in \mathbb{Z}\ \ \text{for all}\ \ 
\tilde M,M \in P^\vee(\mathfrak{g})\ \ 
s.t.\ \ \tilde z\in\tilde H, 
\ \ z\in Z\,. 
\end{equation}
The no-anomaly conditions for subgroups $\tilde H\subset\tilde G$
corresponding to Lie subalgebras $\mathfrak h\subset\mathfrak g$ related 
by inner (outer) automorphisms of $\mathfrak g$ coincide (are simply
related).

\section{Cases with $\mathfrak{h}=\mathfrak{g}$}
\label{sec:3}

As the first step, we shall consider the cases with 
$\mathfrak h=\mathfrak g$ for all simple 
algebras $\mathfrak g$ according to the Cartan classification, and for 
arbitrary nontrivial subgroups $Z\subset\tilde Z$. 
If there are no global gauge anomalies in that case, then the anomalies are 
absent 
also for other $\mathfrak h\subset\mathfrak g$. In other words,
upon restricting $\mathfrak h$ to a smaller subalgebra, the anomalies may 
only disappear. In this way, a lot of trivial cases can be already treated
without specifying the subalgebra $\mathfrak{h}$. We shall then consider 
in the next section the classification of subalgebras $\mathfrak{h}\subset
\mathfrak{g}$ up to conjugation only for the remaining cases: those 
with possible anomalies.

\subsection{Case $A_r = \mathfrak{su}(r+1)$, $r\geq1$}

Lie algebra $\mathfrak{g} = A_r $, corresponding to group 
$\tilde G=SU(r+1)$, is composed of traceless anti-hermitian
matrices of size $r+1$. Its Cartan subalgebra $\mathfrak{t}_{\mathfrak g}$ 
may be taken 
as the subalgebra of diagonal traceless matrices with imaginary entries. 
We define $e_i\in\ii\mathfrak{t}_{\mathfrak g},\ i = 1, \ldots, r+1$, \,as 
a diagonal matrix with the $j$'s diagonal entry equal to $\delta_{ij}$, 
so that tr$(e_i e_j) 
= \delta_{ij}$. Roots (viewed as elements of $\ii\mathfrak{t}_{\mathfrak g}$) 
and coroots of $\mathfrak{su}(r+1)$ have then the form
$e_i - e_j$ for $i \neq j$ and the standard choice of simple roots is 
$\alpha_i = e_i-e_{i+1}, \,i = 1 \ldots r$. The center $\tilde{Z} \cong 
\mathbb{Z}_{r+1}$ may be generated by $z = e^{2i\pi \theta}$ with 
$\theta = \lambda^\vee_r = (1/(r+1)) \sum_{i=1}^{r+1}e_i-e_{r+1}$ where
$\lambda^\vee_i$ denotes the $i$-th simple coweight satisfying
$\,{\rm tr}(\lambda_i^\vee\alpha_j)=\delta_{ij}$. 
Subgroups $Z$ of $\tilde{Z}$ are of the form $Z \cong \mathbb{Z}_p$ with $p | 
(r+1)$, and may be generated by $z^q= e^{2i\pi q \theta} $ 
for $r+1=pq$. The admissible levels
for the WZW model based on group $G=\tilde G/\mathbb Z_p$ are: 
\begin{equation}\label{Consistency_Ar}
\begin{array}{ll}
 k \in 2\mathbb{Z} & \text{if } p \text{ even and } q \text{ odd,}\\
 k \in \mathbb{Z}  & \text{otherwise,}
\end{array}
\end{equation}
see \cite{FGK,Gawedzki}. If we now represent $M$ and $\tilde{M}$ 
in the Euclidian space spanned by vectors $e_i$,
\begin{eqnarray}\label{M_Ar}
&&M = aq \theta = \left( \dfrac{a}{p}, \ldots , \dfrac{a}{p},- \dfrac{ar}{p}
\right), \hspace{2.1cm} a \in \mathbb{Z}\,, \\
\label{tildeM_Ar}
&&\tilde{M} = \tilde{a} \theta = \left( \dfrac{\tilde{a}}{r+1}, \ldots ,
\dfrac{\tilde{a}}{r+1},- \dfrac{\tilde{a}r}{r+1} \right), \qquad \tilde{a} \in
\mathbb{Z}\,,
\end{eqnarray}
the condition for $M$ in (\ref{tzzH}) is satisfied and
$e^{2\ii \pi \tilde{M}}\in \tilde{Z}$. 
\subsubsection{Untwisted case}

If $\omega=I\hspace{-0.03cm}d$, the global gauge invariance
for $\mathfrak{h}=\mathfrak{g}$ is assured if 
\begin{equation}\label{Quantity_Ar}
 k~\text{tr} (M \tilde{M}) = k \dfrac{r a\tilde{a}}{p}\,\in\mathbb Z\,.
\end{equation}
In particular, $k\in p\mathbb{Z}$ is a sufficient condition
for the absence of global anomalies.
Recall that $p$ divides $r+1$. This implies that $p$ and $r$ are relatively 
prime. Hence $k\in p\mathbb{Z}$ is also a necessary condition for the absence
of the anomalies if there are no further restrictions on the values of 
$\tilde a$, \,i.e. if $\mathfrak h=\mathfrak g$. 
Taking into account restrictions (\ref{Consistency_Ar}), this leads to the 
first result:

\begin{prop}\label{prop_Ar_h=g}
The untwisted coset models corresponding to Lie algebra $\mathfrak{g} 
= \mathfrak{su}(r+1)$, subgroups $Z \cong \mathbb{Z}_p$, $r+1 = pq$, 
and arbitrary subalgebras $\mathfrak{h}$ do not have global gauge anomalies 
if $k \in p\mathbb{Z}$. 
The models with $\mathfrak h=\mathfrak g$ and with 
$k \notin p\mathbb{Z}$ for $p>1$ 
odd or $q$ even, or with $k \in 2\mathbb{Z}\setminus p\mathbb{Z}$ 
for $p > 2$ even and $q$ odd 
are anomalous.
\end{prop}

\subsubsection{Twisted case}

For $r>1$, there is one nontrivial outer automorphism of 
$\mathfrak{su}(r+1)$. It maps simple root 
$\alpha_i$ to $\alpha_{r+1-i}$ so that for $\tilde M$ given by 
Eq.\,(\ref{tildeM_Ar}),
\begin{equation}
\omega(\tilde M)
= \omega(\tilde{a} \theta) = \left( \dfrac{\tilde{a}r}{r+1},
\dfrac{-\tilde{a}}{r+1}, \ldots ,\dfrac{-\tilde{a}}{r+1} \right), \qquad 
\tilde{a} \in\mathbb{Z}\,.
\label{omegaM_Ar}
\end{equation}
The condition
\begin{equation}
e^{2\ii\pi\tilde M}\,\omega(e^{-2\ii\pi\tilde M})=e^{4\ii\pi\tilde a\theta}\,\in\,Z
\end{equation}
reduces to the requirement
\begin{eqnarray}
q|\tilde a\quad\ {\rm for}\ \quad q\ \quad {\rm odd\ \quad and}
\quad\ \frac{q}{2}|\tilde a\quad\ {\rm for}\ \quad q\ \ {\rm even}\,.
\end{eqnarray}
It follows that $Z^\omega\cong\mathbb Z_p$ for $\,q\,$ odd and $Z^\omega
\cong\mathbb Z_{2p}$ for $\,q\,$ even.
From Eq.\,(\ref{cc}), we obtain
\begin{eqnarray}
c_{\tilde z\omega(\tilde z)^{-1},z}=\exp\hspace{-0.07cm}
\big[-2\ii\pi k\frac{\tilde a a r}{p}\big]
\end{eqnarray} 
and from Eqs.\,(\ref{M_Ar}) and (\ref{omegaM_Ar}),
\begin{eqnarray}
\exp[-2\ii\pi k\,{\rm tr}(M\omega(\tilde M))]
=\exp\hspace{-0.07cm}\big[-2\ii\pi k\frac{a\tilde a}{p}\big]
\end{eqnarray} 
so that the no-anomaly condition (\ref{idtytw}) reduces to the identity
\begin{equation}
\exp[-2\ii\pi k a\tilde aq]=1
\end{equation}
which always holds implying

\begin{prop}\label{prop_Ar_h=g_tw}
The twisted coset models corresponding to Lie algebra $\mathfrak{g} 
= \mathfrak{su}(r+1)$, subgroups $Z \cong \mathbb{Z}_p$, $r+1 = pq$, 
and arbitrary subalgebras $\mathfrak{h}$ do not have global gauge anomalies.
\end{prop}

\subsection{Case $B_r = \mathfrak{so}(2r+1)$, $r \geq 2$}

Lie algebra $\mathfrak{g} = B_r $, corresponding to group
$\tilde G=Spin(2r+1)$, is composed of real antisymmetric
matrices of size $2r+1$. The Cartan algebra $\mathfrak{t}_{\mathfrak g}$ 
may be taken as composed of $r$ blocks

\begin{equation}
 \begin{pmatrix}
0 & - t_i \\
 t_i  & 0 
 \end{pmatrix}
\end{equation}
placed diagonally, with the last diagonal entry vanishing. Let 
$e_i\in\ii\mathfrak{t}_{\mathfrak g}$ denote the
matrix corresponding to $t_j = \ii\delta_{ij}$. With the normalization such 
that
tr$(e_ie_j) = \delta_{ij}$, roots of $\mathfrak{g}$ have the form $\pm e_i \pm
e_j$ for $i \neq j$ and $\pm e_i$, and one may choose $\alpha_i = e_i -e_{i+1}$
for $i = 1 \ldots r-1$ and $\alpha_r = e_r$ as the simple roots. The center
$\tilde{Z} \cong \mathbb{Z}_2$ is generated by $z = e^{2 \ii \pi \theta}$ with
$\theta = \lambda^\vee_1 = e_1$, and the only nontrivial subgroup of the
center is $Z =
\tilde{Z}$. If we describe $M$ and $\tilde{M}$ in the Euclidian space spanned
by vectors $e_i$, it is enough to take
\begin{equation}
 M = a \theta = \left( a, 0, \ldots, 0 \right), \qquad 
 \tilde{M} = \tilde{a} \theta = \left( \tilde{a}, 0, \ldots, 0 \right), \qquad
a,\tilde{a} \in \mathbb{Z}\,.
\end{equation}
Lie algebra $\mathfrak{so}_{r+1}$ does not have nontrivial
outer automorphisms. For $\omega=I\hspace{-0.03cm}d$, the global gauge 
invariance is assured if
\begin{equation}
 k~\text{tr} (M \tilde{M}) = k a\tilde{a} \in \mathbb{Z}
\end{equation}
which is always the case leading to
\begin{prop}
The coset models corresponding to Lie algebra 
$g = \mathfrak{so}(2r+1)$ and any subalgebra $\mathfrak{h}$ do not have 
global gauge anomalies.
\end{prop}

\subsection{Case $C_r = \mathfrak{sp}(2r)$, $r \geq 3$}

Lie algebra $\mathfrak{g} = C_r$, corresponding to group 
$\tilde G=Sp(2r)$, is composed of antihermitian matrices $X$ of
size $2r$ such that $\Omega X$ is symmetric, with $\Omega$ built 
of $r$ blocks 

\begin{equation}
\omega = \begin{pmatrix}
0 & -1 \\
1 & 0 
 \end{pmatrix}
\end{equation}
placed diagonally. The Cartan algebra $\mathfrak{t}_{\mathfrak g}$ may be taken 
as composed of r blocks $t_i \omega$ placed diagonally. 
Let $e_i\in\ii\mathfrak{t}_{\mathfrak g}$ 
denote the matrix corresponding to
$t_j = \ii\delta_{ij}$. With the normalization tr$(e_ie_j) = 2 \delta_{ij}$, 
roots of $\mathfrak{g}$ have the form $(1/2) (\pm e_i \pm e_j)$ for 
$i \neq j$ and
$\pm e_i$. The simple roots may be chosen as $\alpha_i = (1/2) (e_i - e_{i+1})$
for $i = 1, \ldots r-1$ and $\alpha_r = e_r$. The center $\tilde{Z} \cong 
\mathbb{Z}_2$ is generated by $z = e^{2 \ii \pi \theta}$ with $\theta =
\lambda^\vee_r = (1/2) \sum_{i=1}^{r} e_i$, and its only nontrivial subgroup is
$Z = \tilde{Z}$. We then take $M$ and $\tilde{M}$ in the
Euclidian space spanned by vectors $e_i$ of the form 
\begin{equation}
 M = a \theta = \left( \dfrac{a}{2}, \ldots, \dfrac{a}{2} \right) \qquad 
 \tilde{M} = \tilde{a} \theta = \left( \dfrac{\tilde{a}}{2},  \ldots,
\dfrac{\tilde{a}}{2} \right) \qquad a,\tilde{a} \in \mathbb{Z}.
\end{equation}
 
\noindent Lie algebra $\mathfrak{sp}(2r)$ does not have nontrivial
outer automorphisms. For $\omega=I\hspace{-0.03cm}d$,
taking into account the normalization of $\rm tr$, we obtain: 
\begin{equation}\label{Quantity_Cr}
 k~\text{tr} (M \tilde{M}) = k \dfrac{a\tilde{a}r}{2} ,
\end{equation}
ensuring the global gauge invariance if it is an integer. The admissible 
levels $k$ are 
\begin{eqnarray}
&&k \in \mathbb{Z} \qquad\ \text{\,if } r \text{ is even,}\\
&&k \in 2\mathbb{Z} \qquad \text{if } r \text{ is odd,}
\end{eqnarray}
see \cite{FGK,Gawedzki}, so that the above condition is always satisfied
leading to  

\begin{prop}
The coset models corresponding to Lie algebra 
$\mathfrak{g} = \mathfrak{sp}(2r)$ 
and any subalgebra $\mathfrak{h}$ do not have global gauge anomalies.
\end{prop}

\subsection{Case $D_r = \mathfrak{so}(2r)$, $r \geq 4$}

Lie algebra $\mathfrak{g} = D_r$, corresponding to group 
$\tilde G=Spin(2r)$, is composed of real antisymmetric
matrices of size $2r$. The Cartan algebra $\mathfrak{t}_{\mathfrak g}$
may be taken as composed of $r$ blocks
\begin{equation}
\omega = \begin{pmatrix}
0 & -t_i \\
t_i & 0 
 \end{pmatrix}
\end{equation}
placed diagonally. Let us denote by $e_i\in\ii\mathfrak{t}_{\mathfrak g}$ 
the matrix corresponding to $t_j =\ii\delta_{ij}$. 
With the normalization tr$(e_ie_j) =  \delta_{ij}$, roots of
$\mathfrak{g}$ have the form $\pm e_i \pm e_j$ for $i \neq j$, and the simple
roots may be chosen as $\alpha_i = e_i - e_{i+1}$ for $i = 1 \ldots r-1$ and
$\alpha_r = e_{r-1} + e_{r}$. 

\paragraph{Case of $r$ odd.}

If $r$ is odd, the center $\tilde{Z} \cong \mathbb{Z}_4$ is generated by $z =
e^{2 \ii \pi \theta}$ with $\theta = \lambda^\vee_r = (1/2) \sum_{i=1}^{r} e_i$.
The possible nontrivial subgroups are 
$Z = \tilde{Z}$ and $Z \cong \mathbb{Z}_2$, 
generated by $z^2$. In particular, $Spin(2r)/\mathbb{Z}_2=SO(2r)$. 
Taking the general form of $M$ and $\tilde{M}$ in the Euclidian
space spanned by vectors $e_i$, 
\begin{equation}\label{MMtilde_Dr_odd}
\begin{array} {ll}
 M = a \theta = \left( \dfrac{a}{2}, \ldots, \dfrac{a}{2} \right), & a \in
\mathbb{Z} \text{ if } Z \cong \mathbb{Z}_4\,, \\
 & a \in 2\mathbb{Z} \text{ if } Z \cong \mathbb{Z}_2\,, \\
 \tilde{M} = \tilde{a} \theta = \left( \dfrac{\tilde{a}}{2},  \ldots,
\dfrac{\tilde{a}}{2} \right), & \tilde{a} \in \mathbb{Z}\,.
\end{array}
\end{equation}
The admissibility condition for the levels
in the corresponding WZW models are \cite{Gawedzki}:
\begin{eqnarray}
\label{comp_Dr_odd}
&&k \in 2\mathbb{Z} \text{ if } Z \cong \mathbb{Z}_4\,, \\
&&k \in \mathbb{Z}\hspace{0.06cm}\ \text{ if } Z \cong \mathbb{Z}_2\,.
\end{eqnarray}

\subsubsection{Untwisted case}

If $\omega=I\hspace{-0.03cm}d$ then the global gauge invariance
is assured if the quantity 
\begin{equation}\label{Quantity_Dr_odd}
 k~\text{tr} (M \tilde{M}) = k \dfrac{a\tilde{a}r}{4} ,
\end{equation}
is an integer. The latter holds for 
\begin{eqnarray}
&&k \in 4\mathbb{Z} \text{ if } Z \cong \mathbb{Z}_4\,, \\
&&k \in 2\mathbb{Z} \text{ if } Z \cong \mathbb{Z}_2\,.
\end{eqnarray}

\noindent Comparing to to the admissibility conditions
(\ref{comp_Dr_odd}), we deduce the following 
\begin{prop}\label{prop_Drodd_h=g}
The untwisted coset models corresponding to Lie algebra $\mathfrak{g} =
\mathfrak{so}(2r)$, $r$ odd, and any subalgebra $\mathfrak{h}$ 
do not have global gauge anomalies for
\begin{eqnarray}
&&k \in 4\mathbb{Z} \text{ if } Z \cong \mathbb{Z}_4 \\
&&k \in 2\mathbb{Z} \text{ if } Z \cong \mathbb{Z}_2.
\end{eqnarray}
The models with $\mathfrak h=\mathfrak g$ and $k \in 2\mathbb{Z}$
with odd $k/2$ for $Z \cong \mathbb{Z}_4$
or with $k$ odd for $Z \cong \mathbb{Z}_2$ are anomalous.
\end{prop}

\subsubsection{Twisted case}
There is only one nontrivial outer automorphism $\omega$ of 
$\mathfrak{so}(2r)$ with odd $\,r$, \,It exchanges the simple roots 
$\alpha_{r-1}$ and $\alpha_{r}$ and does not change the other ones. Thus, 
taking $M$ and $\tilde{M}$ given by \eqref{MMtilde_Dr_odd}, we get
\begin{equation}\label{omegaM_Dr_odd}
  \omega(\tilde{M}) = \tilde{a}\,\omega(\lambda_r^\vee) = \tilde{a} 
\lambda_{r-1}^\vee = -\tilde{a}\lambda_r^\vee +\tilde a q^\vee 
= - \tilde M +\tilde a q^\vee
\end{equation}
where $q^\vee \in Q^\vee(D_r)$. The condition
\begin{equation}
e^{2\ii\pi\tilde M}\,\omega(e^{-2\ii\pi\tilde M})=e^{4\ii\pi
\tilde a\theta}\,\in\,Z
\end{equation}
is always satisfied whatever the subgroup $Z \cong \mathbb Z_4$ 
or $\mathbb Z_2$ considered. From Eq.\,(\ref{cc}), we obtain
\begin{eqnarray}
c_{\tilde z\omega(\tilde z)^{-1},z}=\exp\hspace{-0.07cm}\big[
\hspace{-0.05cm}-\ii\pi k\frac{\tilde a a r}{2}\big]
\end{eqnarray} 
and from Eqs.\,(\ref{MMtilde_Dr_odd}) and (\ref{omegaM_Dr_odd}),
\begin{eqnarray}
\exp\hspace{-0.07cm}\big[\hspace{-0.05cm}-2\ii\pi k
\,{\rm tr}(M\omega(\tilde M))\big]
=\exp\hspace{-0.07cm}\big[\hspace{-0.05cm}+\ii\pi k\frac{a\tilde a r}{2}\big]
\end{eqnarray} 
so that the no-anomaly condition (\ref{idtytw}) always holds implying

\begin{prop}\label{prop_Drodd_h=g_tw}
The twisted coset models corresponding to Lie algebra $\mathfrak{g} 
= \mathfrak{so}(2r)$, $r$ odd, subgroups $Z \cong \mathbb{Z}_4$ or 
$\mathbb Z_2$, and arbitrary subalgebras $\mathfrak{h}$ do not have 
global gauge anomalies.
\end{prop}

\paragraph{Case of $r$ even.}

If $r$ is even, the center $\tilde{Z} \cong 
\mathbb{Z}_2 \times \mathbb{Z}_2$ is
generated by $z_1 = e^{2 \ii \pi \theta_1}$ with $\theta_1 = 
\lambda^\vee_r = (1/2)\sum_{i=1}^{r} e_i$
and $z_2 = e^{2 \ii \pi \theta_2}$ with $\theta_2=\lambda^\vee_1 =
e_1$. The possible nontrivial subgroups are given in Table \ref{Subgroups_Dr}.
\begin{table}[htb]
\centering
 \begin{tabular}{|c|c|c|}
\hline
 Subgroup $Z$ & Type & Generator(s) $z_i$ \\
\hline \hline
$\tilde{Z}$ & $\mathbb{Z}_2 \times \mathbb{Z}_2$ & $z_1,\,z_2$ \\
 \hline
$Z_1 := \mathbb{Z}_2 \times \lbrace 1 \rbrace$ & $\mathbb{Z}_2$ & $z_1$ \\
\hline 
$ Z_2 := \lbrace 1 \rbrace \times \mathbb{Z}_2 $ & $\mathbb{Z}_2$  & $z_2$
\\
\hline 
$ Z_{\rm diag}$ & $\mathbb{Z}_2$ & $z_1z_2$ \\
\hline
 \end{tabular}
\caption{Subgroups of $\tilde Z(Spin(2r)) \cong \mathbb{Z}_2 \times
\mathbb{Z}_2$, $r$ even, and their generators. }
\label{Subgroups_Dr}
\end{table}

\noindent Here, $SO(2r)=Spin(2r)/Z_2$. The general form 
of $M$ and $\tilde{M}$ in the Euclidian space spanned
by vectors $e_i$ is
\begin{equation}\label{MMtilde_Dr_even}
\begin{array} {ll}
 M = a_1 \theta_1 + a_2 \theta_2 = \left(\dfrac{a_1}{2} + a_2, \dfrac{a_1}{2},
\ldots, \dfrac{a_1}{2} \right), & a_1,a_2 \in \mathbb{Z} \text{ if } Z =
\tilde{Z}, \\
 & a_1 \in \mathbb{Z}, a_2 = 0 \text{ if } Z = Z_1, \\
 & a_1= 0 , a_2 \in \mathbb{Z} \text{ if } Z = Z_2, \\
 & a_1=a_2 \in \mathbb{Z} \text{ if } Z = Z_{\rm diag}, \\
 \tilde{M} = \tilde{a}_1 \theta_1 + \tilde{a}_2 \theta_2 =
\left(\dfrac{\tilde{a}_1}{2} + \tilde{a}_2, \dfrac{\tilde{a}_1}{2},  \ldots,
\dfrac{\tilde{a}_1}{2} \right), & \tilde{a}_1, \tilde{a}_2 \in \mathbb{Z}.
\end{array}
\end{equation}
In this case, the conditions for admissible levels of the WZW model are 
\cite{Gawedzki}:
\begin{equation}\label{Consistency_Dreven}
\begin{array}{lcl}
 k \in \mathbb{Z} & \text{ if } & r/2 \text{ is even for any } Z, \\
& & r/2 \text{ is odd for } Z = Z_2,\\
 k \in 2\mathbb{Z} & \text{ if } & r/2 \text{ is odd and } Z = \tilde{Z}, Z_1
\text{ or } Z_\text{diag}. \\
\end{array}
\end{equation}

\subsubsection{Untwisted case}

If $\omega=I\hspace{-0.03cm}d$ then the global gauge invariance
is assured if
\begin{equation}\label{Quantity_Dr_even}
 k~\text{tr} (M \tilde{M}) = k \left( \dfrac{a_1 \tilde{a}_1 r}{4} + \dfrac{a_1
\tilde{a}_2}{2} + \dfrac{a_2 \tilde{a}_1}{2}+a_2\tilde a_2 \right) ,
\end{equation}
is an integer. This holds
for $k \in 2\mathbb{Z}$, whatever the subgroup considered. Comparing  
to the admissibility conditions (\ref{Consistency_Dreven}), 
we deduce the following 

\begin{prop}\label{prop_Dreven_h=g}
The untwisted coset models corresponding to Lie algebra $\mathfrak{g} =
\mathfrak{so}(2r)$, $r$ even, and any subalgebra $\mathfrak{h}$ 
do not have global gauge anomalies if $k \in 2\mathbb{Z}$. The models
with $\mathfrak h=\mathfrak g$ and with $k$ odd for $r/2$ even and any 
nontrivial $Z$, or with $k$ odd for 
$r/2$ odd and $Z = Z_2$, are anomalous.
\end{prop}

\subsubsection{Twisted case}
\label{sec:D_reven_tw}

For $r>4$, there is only one nontrivial outer automorphism $\omega$ 
of $\mathfrak{so}(2r)$, which is the same as the one described in 
the case of $r$ odd: it interchanges the simple roots $\alpha_{r-1}$ 
and $\alpha_r$. Thus, taking $M$ and 
$\tilde{M}$ given by \eqref{MMtilde_Dr_even}, we get
\begin{eqnarray}\label{omegaM_Dr_even}
\omega (\tilde{M})&=&\tilde{a}_1\omega(\theta_1) + \tilde{a}_2
\hspace{0.02cm}\omega(\theta_2)\,=\,
\tilde{a}_1 \lambda_{r-1}^\vee + \tilde{a}_2\lambda_1^\vee \cr
&=& \tilde{a}_1\lambda_r^\vee + (\tilde{a}_1+\tilde{a}_2)\lambda_1^\vee + 
\tilde a_1q^\vee\,=\,
\tilde{M} + \tilde{a}_1\theta_2 + \tilde a_1q^\vee
\end{eqnarray}
where $q^\vee \in Q^\vee(D_r)$. The condition
\begin{equation}
e^{2\ii\pi\tilde M}\,\omega(e^{-2\ii\pi\tilde M})
=e^{-2\ii\pi\tilde a_1\theta_2}\,\in\,Z
\end{equation}
is satisfied for arbitrary $\tilde a_1$ if $Z = \tilde Z$ or 
$Z_2$, and for $\tilde a_1=0\,{mod}\,2$ if $Z = Z_1$ or $Z_{diag}$. 
For $Z=\tilde Z$, the expression
for  bihomomorphism (\ref{bihom})
extracted from \cite{FGK} reads:
\begin{eqnarray}
c_{z_1^{m_1}z_2^{m_2},z_1^{n_1}z_2^{n_2}}=\Big(\hspace{-0.1cm}
\pm \exp\hspace{-0.07cm}\big[\dfrac{\ii\pi k}{2}\big]
\Big)^{m_1n_2-m_2n_1}\,
\exp\hspace{-0.07cm}\big[\hspace{-0.05cm}-\frac{\ii\pi k}{2}(m_1n_1 
\dfrac{r}{2}+m_1n_2+m_2n_1+2m_2n_2)\big]
\label{pmbihol}
\end{eqnarray}
for $m_i,n_i\in\mathbb Z$, with the sign $\pm$ corresponding to the two choices
of WZ action functional. For the cyclic subgroups of $\tilde Z$,
the above expression reduces to the one given by Eq.\,(\ref{cc}). 
We have:
\begin{equation}
  c_{\tilde z\omega(\tilde z)^{-1},z}=\big(\pm 1)^{a_1\tilde
    a_1}\,\exp[\ii\pi k( a_1 \tilde a_1 +  a_2 \tilde a_1)]
\end{equation}

\noindent and, from Eqs.\,(\ref{MMtilde_Dr_even}) and (\ref{omegaM_Dr_even}),
\begin{eqnarray}
\exp[-2\ii\pi k\,{\rm tr}(M\omega(\tilde M))]
=\exp\hspace{-0.07cm}\big[\hspace{-0.05cm}-\ii\pi k 
\big( (\dfrac{r}{2}+1) a_1 \tilde a_1 
+ a_1 \tilde a_2 + a_2 \tilde a_1\big)\big].
\end{eqnarray} 
Hence the no-anomaly condition \eqref{idtytw} requires that
\begin{equation}
\label{Dreven_tw_c}
  (\pm 1)^{a_1 \tilde a_1}\exp\hspace{-0.07cm}\big[\hspace{-0.05cm}
- \ii \pi k \big(\dfrac{r}{2} a_1
    \tilde a_1 + a_1 \tilde a_2 \big)\big] = 1
\end{equation}
Considering each subgroup $Z$ and the corresponding values of
$a_1,\,a_2,\,\tilde a_1,$ and $\tilde a_2$, and recalling the
conditions \eqref{Consistency_Dreven} for the admissible levels 
of the corresponding WZW model, we deduce the

\begin{prop}\label{prop_Dreven_h=g_tw}
The twisted coset model corresponding to Lie algebra $\mathfrak{g} 
= \mathfrak{so}(2r)$, $r>4$ even and arbitrary subalgebra 
do not have anomalies for $Z = \tilde Z$ (+ theory), $Z_1$ and
$Z_{diag}$ if $k$ is even, and for $Z = Z_2$ if $k \in \mathbb Z$. 
The twisted models with $\mathfrak h = \mathfrak g$ for 
$Z=\tilde Z$ (- theory) and $k$ even, and for $Z = \tilde Z$, $Z_1$ or 
$Z_{diag}$ and $k$ odd, $r/2>2$ even, are anomalous.
\end{prop}

For $r=4$, there are more nontrivial outer automorphisms, because
the symmetries of the diagram of $D_4$ form the permutation group $S_3$
(the well known ``triality''). They belong to two conjugacy classes, 
the one composed of cyclic permutations of order 2,
\begin{equation}
\omega_1:\,\alpha_3\rightarrow\alpha_4\rightarrow\alpha_3\,,
\qquad \omega_2:\,\alpha_1\rightarrow\alpha_3\rightarrow\alpha_1\,,
\qquad \omega_3:\,\alpha_1\rightarrow\alpha_4\rightarrow\alpha_1\,,
\label{omegas}
\end{equation}
and the one containing cyclic permutations of order 3,
\begin{equation}
\omega_4:\,\alpha_1 \rightarrow \alpha_4 \rightarrow 
\alpha_3\rightarrow \alpha_1\,,\qquad\omega_4^{-1}:\,
\alpha_1 \rightarrow \alpha_3 \rightarrow 
\alpha_4\rightarrow \alpha_1\,.
\label{permutalpha}
\end{equation}
The no-anomaly conditions for twists $\omega$ and 
$\omega'\omega\hspace{0.04cm}\omega'^{-1}$ in the same conjugacy class
are related, as was discussed at the end of Sec.\,\ref{sec:2}:
they coincide if in the latter case subgroup $Z\subset\tilde Z$ is 
replaced by $\omega'(Z)$. The only exception is 
the case $Z=\tilde Z$ for odd $k$ where one has also to interchange  
the $\pm$ theories if $\omega'$ is cyclic of order 2. 
It is straightforward to see that
\begin{equation}
\omega_4\omega_1\omega_4^{-1}=\omega_2\,,\qquad\omega_4^{-1}\omega_1\omega_4=
\omega_3
\end{equation}
and
\begin{equation}
\omega_4(Z_1)=Z_{diag}\,,\qquad\omega_4(Z_2)=Z_1\,,\qquad\omega_4(Z_{diag})
=Z_2\,.
\label{permutZ}
\end{equation}
The results of Proposition \ref{prop_Dreven_h=g_tw} still hold for $r=4$ 
and twist $\omega_1$ and the ones for $r=4$ and twists $\omega_2$ and 
$\omega_3$ follow from the latter by using the above remark 
(or by a direct calculation) giving:  

\begin{prop}\label{prop_D4_h=g_tw}
The twisted coset models corresponding to Lie algebra $\mathfrak{g} 
= \mathfrak{so}(8)$ with twist $\omega_1$ and arbitrary subalgebra 
do not have anomalies for $Z = \tilde Z$ (+ theory), $Z_1$ and
$Z_{diag}$ if $k$ is even, and for $Z = Z_2$ if $k \in \mathbb Z$. 
The models with $\mathfrak h = \mathfrak g$ for $Z=\tilde Z$ (- theory)
and $k$ even, and for $Z = \tilde Z$, $Z_1$ or $Z_{diag}$ and $k$ odd are 
anomalous. The results for twist $\omega_2$ ($\omega_3$) are as 
the ones for twist $\omega_1$ except for the permutation (\ref{permutZ})
of the subgroups $Z\rightarrow\omega_4(Z)$ ($Z\rightarrow\omega_4^{-1}(Z))$.
\end{prop}

\noindent For the cyclic outer automorphism $\omega_4$ of order 3,
taking $M$ and $\tilde{M}$ given 
by Eqs.\,\eqref{MMtilde_Dr_even}, we obtain:
\begin{equation}\label{omega2M_Dr_even}
  \omega_4 (\tilde{M}) =
 (\tilde{a}_1 + \tilde{a}_2) \theta_1 + \tilde{a}_1 
\theta_2 + \tilde a_1 q^\vee
\end{equation}
where $q^\vee \in Q^\vee(D_4)$. The condition
\begin{equation}
e^{2\ii\pi\tilde M}\,\omega_4(e^{-2\ii\pi\tilde M})=\exp[2\ii\pi(-\tilde
  a_2\theta_1+ (\tilde a_2 - \tilde a_1) \theta_2)]\,\in\,Z
\end{equation}
is satisfied for arbitrary $\tilde a_1, \tilde a_2$ if $Z = \tilde Z$,
and for $\tilde a_1 =\tilde a_2$, $\tilde a_2 = 0$, $\tilde a_1 = 0$,
all $mod\,2$, if $Z = Z_1,\, Z_2$ or $Z_{diag}$ respectively. 
Expression (\ref{pmbihol}) for the bihomomorphism gives here:
\begin{equation}
   c_{\tilde z\hspace{0.02cm}\omega_4(\tilde z)^{-1},z}
= (\pm 1)^{-a_2 \tilde a_2 + a_1
     \tilde a_1 - a_1 \tilde a_2}\exp[\ii \pi k(a_1\tilde a_1+ a_2 
\tilde a_1 - a_2 \tilde a_2)]
\end{equation}
From Eqs.\,(\ref{MMtilde_Dr_even}) and (\ref{omegaM_Dr_even}),
\begin{eqnarray}
\exp[-2\ii\pi k\,{\rm tr}(M\omega_4(\tilde M))]
=\exp\hspace{-0.07cm}\big[\hspace{-0.05cm}-\ii\pi k 
\big( a_1 \tilde a_1 + a_2 \tilde a_1 + a_2 \tilde
  a_2 \big)\big]
\end{eqnarray} 
so that the no-anomaly condition \eqref{idtytw} becomes
\begin{equation}
  (\pm 1)^{-a_2 \tilde a_2 - a_1
     \tilde a_2 + a_1 \tilde a_1}= 1.
\end{equation}
Considering each subgroup $Z$ and the corresponding values of
$a_1,\,a_2,\,\tilde a_1,$ and $\tilde a_2$, and recalling the
admissible values \eqref{Consistency_Dreven} of the level, we deduce

\begin{prop}\label{prop_D4_h=g_tw2}
The twisted coset models corresponding to Lie algebra $\mathfrak{g} 
= \mathfrak{so}(8)$, outer automorphism $\omega_4$ and arbitrary
subalgebra do not have anomalies for $Z=\tilde Z$ (+ theory) and
$Z=Z_1,\,Z_2$ or $Z_{diag}$. The models with $\mathfrak h = \mathfrak g$ 
and $Z = \tilde Z$ (- theory) is anomalous.  
\end{prop}

\noindent The results for the twist $\omega_4^{-1}$ may be deduced from
the above proposition if we observe that $\omega_4^{-1}$ may be obtained
from $\omega_4$ by the conjugation by any cyclic outer automorphism
$\omega'$ of order 2. Hence the conditions for the absence or the presence
of anomalies for the theory twisted by $\omega_4^{-1}$ are as for
the ones for the twist $\omega_4$ except for the exchange of the $\pm$
theories for $Z=\tilde Z$ and $k$ odd leading to

\begin{prop}\label{prop_D4_h=g_tw3}
The twisted coset models corresponding to Lie algebra $\mathfrak{g} 
= \mathfrak{so}(8)$, outer automorphism $\omega_4^{-1}$ and arbitrary
subalgebra do not have anomalies for $Z=\tilde Z$ (($-$)${}^k$ theory) and
$Z=Z_1,\,Z_2$ or $Z_{diag}$. The models with $\mathfrak h = \mathfrak g$, 
and $Z = \tilde Z$ (($-$)${}^{k+1}$ theory) is anomalous.  
\end{prop}

\noindent This may be confirmed by a direct calculation.

\subsection{Case $\mathfrak{e}_6$}

The imaginary part $\,\ii\mathfrak{t}_{\mathfrak g}\,$ of the 
complexification of the Cartan 
subalgebra $\mathfrak{t}_{\mathfrak g}$ of $\mathfrak{g} = \mathfrak{e}_6$ 
may be identified 
with the subspace of $\mathbb{R}^{7}$ orthogonal to the vector 
$(1, \ldots, 1, 0)$, with the scalar product inherited from 
$\mathbb{R}^{7}$. The simple roots may be
taken as $\alpha_i = e_i - e_{i+1}$ for $i=1 \ldots 5$ and $\alpha_6 =
(1/2)(-e_1-e_2-e_3+e_4+e_5+e_6) + (1/\sqrt{2})e_7$, where $e_i$ are the vectors
of the canonical basis of $\mathbb{R}^{7}$. The center $\tilde{Z} \cong
\mathbb{Z}_3$ is generated by $z = e^{2 \ii \pi \theta}$ with $\theta =
\lambda^\vee_5 = (1/6)(e_1 +e_2+e_3+e_4+e_5-5e_6) + (1/\sqrt{2})e_7$. The only
nontrivial subgroup is $Z = \tilde{Z}$. The general form of $M$ and
$\tilde{M}$ in the Euclidian space spanned by vectors $e_i$ is

\begin{equation}\label{MMtilde_e6}
\begin{array}{l}
 M = a \theta = \left( \dfrac{a}{6},\ldots ,\dfrac{a}{6},\dfrac{-5a}{6},
\dfrac{a}{\sqrt{2}} \right) \qquad  a \in \mathbb{Z},\\
 \tilde{M} = \tilde{a} \theta = \left( \dfrac{\tilde{a}}{6},\ldots
,\dfrac{\tilde{a}}{6},\dfrac{-5\tilde{a}}{6}, \dfrac{\tilde{a}}{\sqrt{2}}
\right) \qquad \tilde{a} \in \mathbb{Z}.
\end{array}
\end{equation}

\subsubsection{Untwisted case}

If $\omega=I\hspace{-0.03cm}d$ then the global gauge invariance
is assured if
\begin{equation}\label{Quantity_e6}
 k~\text{tr} (M \tilde{M}) = k \dfrac{4a\tilde{a}}{3} ,
\end{equation}
is an integer. This holds
for $k \in 3\mathbb{Z}$. Since all integer levels $k \in \mathbb{Z}$ 
are admissible \cite{FGK,Gawedzki}, 
we deduce 

\begin{prop}
The untwisted coset models corresponding to Lie algebra $\mathfrak{g} =
\mathfrak{e}_6$ and arbitrary subalgebra $\mathfrak{h}$ do not have global 
gauge anomalies if $k \in 3\mathbb{Z}$. The models $Z=\mathbb Z_3$, 
$\mathfrak h=\mathfrak g$ and $k \in
\mathbb{Z} \setminus 3\mathbb{Z}$ are anomalous.
\end{prop}

\subsubsection{Twisted case}
There is only one nontrivial outer automorphism $\omega$ of 
$\mathfrak{e}_6$, which exchanges the simple roots $\alpha_1$ 
and $\alpha_2$ with $\alpha_5$ and $\alpha_4$ and does not change 
the other ones. Thus, taking $M$ and $\tilde{M}$ given by 
\eqref{MMtilde_e6}, we get
\begin{equation}\label{omegaM_e6}
  \omega(\tilde{M}) = \tilde{a}\hspace{0.02cm}\omega(\lambda_5^\vee) 
= \tilde{a} \lambda_{1}^\vee = -\tilde{a}\lambda_5^\vee + \tilde a q^\vee 
= - \tilde M + \tilde aq^\vee
\end{equation}
where $q^\vee \in Q^\vee(\mathfrak e_6)$. The condition
\begin{equation}
e^{2\ii\pi\tilde M}\,\omega(e^{-2\ii\pi\tilde M})=e^{4\ii\pi\tilde a\theta}\,\in\,Z
\end{equation}
is always satisfied for $Z = \tilde{Z}$. From Eq.\,(\ref{cc}), we obtain
\begin{eqnarray}
c_{\tilde z\omega(\tilde z)^{-1},z}=\exp[-2\ii\pi k\frac{4 \tilde a a}{3}]
\end{eqnarray} 
and from Eqs.\,(\ref{MMtilde_e6}) and (\ref{omegaM_e6}),
\begin{eqnarray}
\exp[-2\ii\pi k\,{\rm tr}(M\omega(\tilde M))]
=\exp\hspace{-0.07cm}\big[\hspace{-0.05cm}+2\ii\pi k\frac{4 a\tilde a }{3}
\big]
\end{eqnarray} 
so that the no-anomaly condition (\ref{idtytw}) always holds implying

\begin{prop}\label{prop_e6_h=g_tw}
The twisted coset models corresponding to Lie algebra $\mathfrak{g} 
= \mathfrak e_6$, subgroup $Z \cong \mathbb{Z}_3$ and arbitrary subalgebras 
$\mathfrak{h}$ do not have global gauge anomalies.
\end{prop}

\subsection{Case $\mathfrak{e}_7$}

The imaginary part $\,\ii\mathfrak{t}_{\mathfrak g}\,$ of the 
complexification of the Cartan 
subalgebra $\mathfrak{t}_{\mathfrak g}$ of $\mathfrak{g} = \mathfrak{e}_7$ 
may be 
identified with the
subspace of $\mathbb{R}^{8}$ orthogonal to the vector $(1, \ldots, 1)$ with the
simple roots $\alpha_i = e_i - e_{i+1}$ for $i=1 \ldots 6$ and $\alpha_7 =
(1/2)(-e_1-e_2-e_3-e_4+e_5+e_6+e_7+e_8)$, where $e_i$ are the vectors of the
canonical basis of $\mathbb{R}^{8}$. The center $\tilde{Z} \cong \mathbb{Z}_2$
is generated by $z = e^{2 \ii \pi \theta}$ with $\theta = \lambda^\vee_1 =
(1/4)(3,-1,\ldots, -1, 3)$. The only nontrivial subgroup is $Z = \tilde{Z}$.
The general form of $M$ and $\tilde{M}$ in the Euclidian space generated
by $e_i$ is 
\begin{equation}
\begin{array}{l}
 M = a \theta = \left( \dfrac{3a}{4},\dfrac{-a}{4}, \ldots, \dfrac{-a}{4},
\dfrac{3a}{4} \right) \qquad  a \in \mathbb{Z},\\
 \tilde{M} = \tilde{a} \theta = \left( \dfrac{3\tilde{a}}{4},
\dfrac{-\tilde{a}}{4}, \ldots,\dfrac{-\tilde{a}}{4}, \dfrac{3\tilde{a}}{4}
\right) \qquad \tilde{a} \in \mathbb{Z}.
\end{array}
\end{equation}
Lie algebra $\mathfrak{e}_7$ does not have nontrivial outer automorphisms
so that we may take $\omega=I\hspace{-0.03cm}d$.
The global gauge invariance is then assured if the quantity
\begin{equation}\label{Quantity_e7}
 k~\text{tr} (M \tilde{M}) = k \dfrac{3a\tilde{a}}{2} ,
\end{equation}
is an integer. This holds
for $k \in 2\mathbb{Z}$. The condition for admissible levels 
also requires in this case that $k \in 2\mathbb{Z}$ \cite{FGK,Gawedzki} 
so that we deduce: 

\begin{prop}
The coset models corresponding to Lie algebra $\mathfrak{g} =
\mathfrak{e}_7$ and any subalgebra $\mathfrak{h}$ do not have global
gauge anomalies.
\end{prop}

\subsection{Case $\mathfrak{g}_2$, $\mathfrak{f}_4$ and $\mathfrak{e}_8$}

The center of the simply connected groups  corresponding to Lie
algebras  $\mathfrak{g} =
\mathfrak{g}_2, \mathfrak{f}_4$ or $\mathfrak{e}_8$ is trivial~: 
$\tilde{Z} \cong \lbrace 1\rbrace$ so that there are no nontrivial
subgroups $Z$ in that case and we infer:

\begin{prop}
The coset models corresponding to Lie algebras $\mathfrak{g} =
\mathfrak{g}_2, \mathfrak{f}_4$ or $\mathfrak{e}_8$ and any subalgebra
$\mathfrak{h}$ do not have global gauge anomalies.
\end{prop}

\section{Regular subalgebras}
\label{sec:4}

Looking back at the previous section, the global gauge anomalies of the coset
models may appear only for $\mathfrak{g} = A_r$, $D_r$ 
and $\mathfrak{e}_6$ in the untwisted case, and only for $\mathfrak{g}=D_r$
with even $r$ in the twisted case (note that these are all simply laced 
Lie algebras). Now we have to specify the Lie subalgebra $\mathfrak{h}$ of 
a simple algebra $\mathfrak g $ to see in which cases the anomalies 
survive the restriction of the symmetry group. The first class of 
semisimple subalgebras that we shall consider are the regular ones, 
introduced by Dynkin in \cite{Dynkin}. A Lie subalgebra $\mathfrak{h}$ of 
an algebra $\mathfrak{g}$ is called regular if, for a choice of the Cartan 
subalgebra $\,t_{\mathfrak g}\subset\mathfrak g\,$
(defined up to conjugation), it's complexification is of the form
\begin{equation}
\mathfrak h^{\mathbb C}\,=\,\mathfrak t_{\mathfrak
h}^{\mathbb C}\oplus\Big(\mathop{\oplus}\limits_{\alpha\in
\Delta_{\mathfrak h}\subset\Delta_{\mathfrak g}}\mathbb C e_\alpha\Big)
\end{equation}
where $\mathfrak t_{\mathfrak h}\subset\mathfrak t_{\mathfrak g}$ is
a Cartan subalgebra of $\mathfrak{h}$. Subalgebra $\,\mathfrak h\,$ 
is semisimple if $\,\alpha\in\Delta_{\mathfrak h}\,$ implies that 
$\,-\alpha\in\Delta_{\mathfrak h}\,$ and if 
$\,\alpha\in\Delta_{\mathfrak h}\,$  span $\,\mathfrak t_{\mathfrak h}^{\mathbb C}$.
$\,\Delta_{\mathfrak h}\,$ is then the set of roots of $\,\mathfrak h$. 

\paragraph{Construction of regular subalgebras.}
There is a nice diagrammatic method to obtain all the regular 
semisimple subalgebras 
of a given semisimple algebra (up to conjugation), 
proposed by Dynkin in \cite{Dynkin} and summarized in \cite{Lorente}. 
We briefly describe it here:

\begin{enumerate}
 \item Take the Dynkin diagram of the ambient algebra $\mathfrak{g}$, 
and adjoin to it a node corresponding to the lowest root $\delta=-\phi$ 
(negative of the highest root $\phi$) of $\mathfrak{g}$, obtaining the 
extended Dynkin diagram of $\mathfrak{g}$.

\item Remove arbitrarily one root from this diagram, in order to obtain 
at most $r+1$ different diagrams, which may split into orthogonal subdiagrams.
\item Reapply the firsts two steps to each connected subdiagram obtained
above, until no new diagram appears. This way one gets all the regular 
subalgebras $\mathfrak{h} \subset \mathfrak{g}$ of maximal rank.
\item Remove again an arbitrarily root from each diagram, and apply the full 
procedure to each connected subdiagram obtained this way (including the
last step).
\end{enumerate}
The algorithm stops when no root can be removed, hence one will obtain 
all the regular subalgebras of $\mathfrak{g}$.

\subsection{Regular semisimple subalgebras of $A_r$}

The semisimple regular subalgebras of $A_r$ are 
given in \cite{Dynkin} (Chapter II, Table 9) and have the form: 

\begin{equation}\label{regular_subalg_Ar}
 \mathfrak{h} = A_{r_1} \oplus \ldots \oplus A_{r_m}, \qquad r_1 + 1 + \ldots +
r_m + 1 \leq r+1
\end{equation}

\noindent The embedding of $\mathfrak{h}$ in $\mathfrak{g}$ realizing
the ideals $A_{r_i}$ as diagonal blocks in the matrices of $A_{r}$
is unique up to an inner automorphism of $A_r$.
Taking $M$ and $\tilde{M}$ as given in Eqs. \eqref{M_Ar} and \eqref{tildeM_Ar} 
we must require that $\tilde M+q^\vee\in\ii\mathfrak{t}_{\mathfrak h}$, 
for some $q^\vee\in Q^\vee(A_r)$. Looking block by block, we obtain 
the conditions 
\begin{equation}
\label{condA}
 \dfrac{\tilde{a}(r_i+1)}{r+1} \in \mathbb{Z} \qquad \forall i = 1, \ldots, m
\end{equation}
and that
\begin{equation}
\dfrac{\tilde{a}}{r+1}\in\mathbb Z
\end{equation}
if the inequality in (\ref{regular_subalg_Ar}) is strict. The latter condition
implies that (\ref{Quantity_Ar}) holds eliminating possible global 
gauge anomalies. We may then limit ourselves to the case when the inequality 
in (\ref{regular_subalg_Ar}) is saturated. this implies that 
For $i = 1, \ldots, m$, we may then rewrite conditions (\ref{condA}) as 
\begin{equation}
 \tilde{a}(r_i+1) = q_i (r+1) \qquad q_i \in \mathbb{Z}.
\label{lst}
\end{equation}
In what follows, we shall denote by, respectively, 
$u_1 \wedge\cdots\wedge u_n$ and $u_1\vee\cdots\vee u_n$ the greatest 
common divisor and the least common multiple of $u_1,\dots,u_n$. 
Dividing both sides of Eq.\,(\ref{lst}) by $(r+1) \wedge (r_i+1)$, we get
\begin{equation}
 \tilde{a}\frac{r_i+1}{(r+1) \wedge (r_i+1)}= q_i\frac{r+1}{(r+1) 
\wedge (r_i+1)}
\end{equation}
so that $\frac{r+1}{(r+1) 
\wedge (r_i+1)}|\frac{\tilde{a}\,(r_i+1)}{(r+1)\wedge(r_i+1)}$. Using the 
fact that $\frac{r+1}{(r+1)\wedge (r_i+1)}$ and 
$\frac{r_i+1}{(r+1)\wedge(r_i+1)}$ are relatively prime, we infer 
that $\frac{r+1}{(r+1) 
\wedge (r_i+1)}|\tilde a$, i.e. that
\begin{equation}
 \tilde{a} \in \dfrac{r+1}{(r+1) \wedge (r_i+1)}\mathbb{Z} \qquad \forall i
= 1, \ldots, m
\end{equation}
which leads, according to Proposition \ref{lcm} of Appendix \ref{app:2}, 
to the condition
\begin{equation}
 \tilde{a} \in \left(\dfrac{r+1}{(r+1) \wedge (r_1+1)} \vee \cdots \vee
\dfrac{r+1}{(r+1) \wedge (r_m+1)} \right) \mathbb{Z}
\end{equation}
This property can be reformulated, using Proposition
\ref{lcmfrak} of Appendix \ref{app:2}, as
\begin{equation}\label{Condition_regular_Ar}
 \tilde{a} \in \left(\dfrac{r+1}{(r+1) \wedge (r_1+1) \wedge \cdots \wedge
(r_m+1)} \right) \mathbb{Z}
\end{equation}
Since we assumed that $r_1 + 1 + \ldots + r_m + 1 = r+1$,
condition (\ref{Condition_regular_Ar}) may be simplified to 
\begin{equation}
 \tilde{a} \in \left(\dfrac{r+1}{(r_1+1) \wedge \cdots \wedge (r_m+1)} \right)
\mathbb{Z}
\end{equation}
In order to guarantee that the quantity \eqref{Quantity_Ar}
is an integer for every $a$ and $\tilde{a}$, ensuring the global 
gauge invariance, it is enough to compute it for $a = 1$ and
\begin{equation}
 \tilde{a} = \dfrac{r+1}{(r_1+1) \wedge \cdots \wedge (r_m+1)}\,.
\end{equation}
Denoting $(r_1+1) \wedge \ldots \wedge
(r_m+1)=l$, and $r+1 = pq$, the quantity \eqref{Quantity_Ar} becomes
\begin{equation}
 k~\text{tr} (M \tilde{M}) = k \dfrac{r q}{{l}} = k r\dfrac{q/(q\wedge
{l})}{{l}/(q\wedge {l})}.
\label{lst1}
\end{equation}
Finally, recalling that ${l} | (r+1)$ and, consequently, 
$\frac{l}{q\wedge {l}}$ and $r$ are relatively prime, 
we infer that the right hand side of Eq.\,(\ref{lst1})
is be an integer if and only if 
\begin{equation}
 k \in \dfrac{{l}}{q \wedge {l}} \mathbb{Z}\,.
\end{equation}
Taking into account condition \eqref{Consistency_Ar}  for admissible
levels, we are now able to state 
\begin{prop}
The untwisted coset models built with Lie algebra $\mathfrak{g} = A_r$, 
subgroup $Z \cong \mathbb{Z}_p$ for $ (r+1)=pq$ and any regular subalgebra
$\mathfrak{h} = A_{r_1} \oplus \ldots \oplus A_{r_m}$ do not have global 
gauge anomalies for 
\begin{itemize}
 \item $r_1+1 + \ldots r_m+1 < r+1 \qquad k \in 
\left\lbrace\begin{array}{l}
2\mathbb{Z} \text{ if } p \text{ even and } q \text{ odd}\\ 
\mathbb{Z} \text{ otherwise}\end{array}\right.$

\item $r_1+1 + \ldots r_m+1 = r+1 \qquad k \in 
\left\lbrace\begin{array}{l}
\dfrac{{l}}{q \wedge {l}} \mathbb{Z} \cap 2\mathbb{Z} \text{ if } p \text{ even
and } q \text{ odd} \\
\dfrac{{l}}{q \wedge {l}} \mathbb{Z} \text{ otherwise}
\end{array}\right.$
\end{itemize}
where ${l} = (r_1+1) \wedge \ldots \wedge (r_m+1)$.
The other untwisted models with admissible levels are anomalous.
\end{prop}

\paragraph{Example 1:  $\mathfrak g = A_4 = \mathfrak{su}(5)$.}

The center $\tilde Z \cong \mathbb Z_5$ of the corresponding group has 
only one nontrivial subgroup, $Z = \tilde Z \cong \mathbb Z_5$, so with 
$p=5$ odd and $q =1$ odd with the previous notations. The admissible 
levels are $k \in \mathbb Z$, according 
to \eqref{Consistency_Ar}. 
Following Proposition \ref{prop_Ar_h=g}, the regular subalgebra 
$\mathfrak h = \mathfrak g$ leads to the condition $k \in 5\mathbb Z$ 
for non-anomalous models. Then, applying the last proposition above, 
the cases $\mathfrak h = A_1,\,A_1 \oplus A_1\equiv 2A_1,\,A_2$ and $A_3$ 
leads to non-anomalous models for every $k \in \mathbb Z$, because 
here we have 
$r_1+1 + \ldots r_m+1 < r+1 = 5$. For $\mathfrak h = A_2 \oplus A_1$, 
we have an equality. However, $l = (r_1+1) \wedge (r_2+1) = 3 \wedge 2 = 1$, 
so $l / (l \wedge q) = 1$ and the model has no anomalies 
for every $k \in \mathbb Z$. Consequently, the only anomalous models 
corresponding to $\mathfrak g = A_4$ and $\mathfrak h$ regular are those 
with $\mathfrak h = \mathfrak g$, $Z = \tilde Z$ and $k \in \mathbb Z 
\setminus 5\mathbb Z$.

\paragraph{Example 2:  $\mathfrak g = A_5 = \mathfrak{su}(6)$.}

Here the center $\tilde Z \cong \mathbb Z_6$ has three nontrivial 
subgroups : $Z \cong \mathbb Z_6, \mathbb Z_3$ and $\mathbb Z_2$ with 
the respective admissible levels $k \in 2 \mathbb Z,\,\mathbb Z$ and 
$2 \mathbb Z$. The models corresponding to the case 
$\mathfrak h = \mathfrak g$ will be non-anomalous for 
\begin{equation}
  k \in \left\lbrace
  \begin{array}{ll}
    6 \mathbb Z & \text{if } Z \cong \mathbb Z_6 \\
    3 \mathbb Z & \text{if } Z \cong \mathbb Z_3 \\    
    2 \mathbb Z & \text{if } Z \cong \mathbb Z_2.
  \end{array}
\right.
\end{equation}
Regular subalgebras $\mathfrak h = A_1,\,2A_1,\,A_2,\,A_2 \oplus A_1, 
A_3$ and $A_4$ correspond to the strict inequality for ranks in 
the proposition above, so there will be no anomalies for these models with
\begin{equation}
  k \in \left\lbrace
  \begin{array}{ll}
    2 \mathbb Z & \text{if } Z \cong \mathbb Z_6 \text{ or } \mathbb Z_2 \\
    \mathbb Z & \text{if } Z \cong \mathbb Z_3. \\    
  \end{array}
\right.
\end{equation}
Computation shows that $\mathfrak h = 2A_2$ leads to non-anomalous 
models for the same $k$ as for $\mathfrak h = \mathfrak g$, and that 
the models corresponding to $\mathfrak h = A_3 \oplus A_1$ and to $3A_1$ 
have no anomalies for $k \in 2 \mathbb Z$ if 
$Z \cong \mathbb Z_6$ or $\mathbb Z_2$ and for $k\in\mathbb Z$ 
if $Z\cong\mathbb Z_3$ . 
Thus, the anomalous models corresponding to $\mathfrak g = A_5$ have either 
$\mathfrak h = \mathfrak g$ or $\mathfrak{h}=2A_2$, 
where $k \in 2\mathbb Z \setminus 6 \mathbb Z$ for $Z \cong \mathbb Z_6$ 
and $k \in \mathbb Z \setminus 3 \mathbb Z$ for $Z \cong \mathbb Z_3$. 

\subsection{Regular semisimple subalgebras of $D_r$}

The semisimple regular subalgebras of $D_r$ are given 
in \cite{Dynkin} (Chapter II, Table 9) and have the form:

\begin{equation}\label{regular_subalg_Dr}
 \mathfrak{h} = A_{r_1} \oplus \ldots \oplus A_{r_m} \oplus D_{s_1} \oplus
\ldots \oplus D_{s_n}
\end{equation}
where $r_1 + 1 + \ldots + r_m + 1 + s_1 + \ldots s_n \leq r$.\footnote{To 
take into account all the possible cases with this formula, we may need 
to consider $D_2$ instead of $2A_1$ and $D_3$ instead of $A_3$ 
to respect the inequality. See examples below.} The embedding of
$D_{s_i}$ subalgebras realizes them as diagonal 
blocks in $D_r$. Instead of giving an explicit embedding of 
subalgebras $A_{r_i}$, it is enough to see that
$A_l$ is trivially embedded in $D_{l+1}$, by sending the $l$ simple roots
$\alpha_i^{A_l}$ of $A_l$ to the $l$ first simple roots $\alpha_i^{D_{l+1}}$ of
$D_{l+1}$. Then, the Serre construction allows us to reconstruct the full
structure of $A_l$, embedded in $D_{l+1}$, which is then easily embedded in
$D_r$ as a diagonal block. The embedding of $\mathfrak h$ into $\mathfrak g$ 
described above is unique, up to inner automorphisms of $\mathfrak g$, except
for even $r$ if there are no $D_{s_i}$ and $r_1+1+\ldots+r_m+1=r$ with 
all $r_i$ odd. In the latter case there is a second 
independent embedding of $A_{r_1}\oplus\ldots\oplus A_{r_m}$ into $D_r$ that
sends the simple roots of $A_{r_m}$ to the last $r_m+1$ simple roots of 
$D_r$ omitting $\alpha_{r-1}$. That embedding is related to the previous 
one by the outer automorphism $\omega$ of $D_r$ that permutes roots 
$\alpha_{r_1}$ and $\alpha_r$, but not by an inner automorphism. 
Recall that the coroot lattice $Q^\vee(D_r)$ is composed of vectors 
\begin{equation}
q^\vee=\sum\limits_{i=1}^rq^\vee_ie_i\quad{\rm with}\quad 
q^\vee_i\in\mathbb Z\quad{\rm and}\quad\sum\limits_{i=1}^rq^\vee_i\in2\mathbb Z\,.
\label{coroot_lat_Dr}
\end{equation}

\paragraph{Case of $r$ odd.}
Taking $M$ and $\tilde{M}$ as given in \eqref{MMtilde_Dr_odd}, 
we shall impose the condition $e^{2\ii \pi \tilde{M}}\in \tilde{H}$. On the 
Lie-algebra level, we have to show that for some 
$q^\vee\in Q^\vee(\mathfrak{g})$, $\tilde M+q^\vee$ belongs to
$\ii\mathfrak{t}_{\mathfrak{h}}$. Looking block by block, we infer that 
\begin{equation}
\label{D1}
\frac{\tilde{a}\hspace{0.01cm}(r_i+1)}{2}\in\mathbb Z\,, \qquad i=1, \dots, m\,,
\end{equation}
and that
\begin{equation}
\label{D2}
\frac{\tilde{a}}{2}\in\mathbb Z
\end{equation}
if $r_1 + 1 + \ldots + r_m + 1 + s_1 +\ldots s_n < r$. The condition 
that the sum of components of vectors in $Q^\vee(\mathfrak{so}(2r))$ is even
imposes the additional requirement that   
\begin{equation}
\label{add_odd}
\frac{\tilde{a}\hspace{0.01cm}r}{2}\in2\mathbb Z\,,
\end{equation}
i.e. $\tilde a\in 4\mathbb Z$, in the absence of $D_{s_i}$ components 
in $\mathfrak{h}$,  (in that case conditions (\ref{D1}) and (\ref{D2}) imply 
already that $\tilde a\in2\mathbb Z$).
Re-examining the quantity \eqref{Quantity_Dr_odd} which 
has to be an integer with the above restrictions in mind 
and taking into account the conditions for admissible levels, 
we deduce 

\begin{prop}
The untwisted coset models built with Lie algebra $\mathfrak{g} =
\mathfrak{so}(2r)$, $r$ odd, and a regular subalgebra  $\mathfrak{h} = A_{r_1}
\oplus \ldots \oplus A_{r_m} \oplus D_{s_1} \oplus \ldots \oplus D_{s_n}$ 
do not have global gauge anomalies for the following cases 
\begin{itemize}

\item\hspace{-0.1cm}$r_1 + 1 + \ldots + r_m + 1 + s_1 + \ldots s_n = r \text{ with all }
r_i \text{ odd and }\quad k \in \left\lbrace
\begin{array}{l} 
4\mathbb{Z} \text{ if } Z \cong \mathbb{Z}_4 \\
 2\mathbb{Z} \text{ if } Z \cong \mathbb{Z}_2\,
\end{array}\right.$
\item $\hspace{-0.3cm}\left.\begin{array}{l}r_1 + 1 + \ldots + r_m + 1 + s_1 + \ldots s_n < r 
\text{ or}\\ r_1 + 1 + \ldots + r_m + 1 + s_1 + \ldots s_n = r\ and\ some\  
r_{i}\ even\end{array}\right\rbrace
\ k \in
\left\lbrace
\begin{array}{l} 
2\mathbb{Z} \text{ if } Z \cong \mathbb{Z}_4 \\
\mathbb{Z}\ \hspace{0.04cm} \text{ if } Z \cong \mathbb{Z}_2
\end{array}\right.$
\end{itemize}
The other untwisted models with admissible levels 
are not globally gauge invariant.
\end{prop}

\noindent{\bf Remark}\ \ \ In particular, the global gauge anomalies present
if $\mathfrak h=\mathfrak g$ for $Z=\mathbb Z_4$ and $k\in2\mathbb Z$,
$k/2$ odd, or for $Z=\mathbb Z_2$ and $k$ odd, disappear for $\mathfrak h= 
A_{r_1}\oplus \ldots\oplus A_{r_m}\oplus D_{s_1}\oplus\ldots\oplus D_{s_n}$
if $r_1 + 1 + \ldots + r_m + 1 + s_1 + \ldots s_n < r$ or if
$r_1 + 1 + \ldots + r_m + 1 + s_1 + \ldots s_n = r$ with some
$r_i$ even. Note that if there no $D_{s_i}$ and $r_1 + 1 + \ldots + r_m + 1=r$
then all $r_i$ cannot be odd.

\paragraph{Example: $\mathfrak{g}=D_5=\mathfrak{so}(10)$.} The admissible
levels are $k \in 2\mathbb Z$ for 
$Z = \tilde Z \cong \mathbb Z_4$ and $k \in \mathbb Z$ for $Z \cong 
\mathbb Z_2$. According to Proposition \ref{prop_Drodd_h=g}, there are 
no gauge anomalies in the case $\mathfrak h = \mathfrak g$ for 
\begin{equation}\label{ex_Drodd_1}
  k \in \left\lbrace
    \begin{array}{ll}
      4 \mathbb Z \text{ if } Z \cong \mathbb Z_4\\
      2 \mathbb Z \text{ if } Z \cong \mathbb Z_2.\\
    \end{array}
\right.
\end{equation}
 For regular subalgebra $\mathfrak h = A_1, 2A_1\cong D_2, 
A_2, A_3\cong D_3, D_4$, 
the inequality on the ranks is strict so there are no anomalies for  
\begin{equation}\label{ex_Drodd_2}
  k \in \left\lbrace
  \begin{array}{ll}
    2 \mathbb Z & \text{if } Z = \mathbb Z_4 \\
    \mathbb Z & \text{if } Z = \mathbb Z_2. \\    
  \end{array}
\right.
\end{equation}
In the case $\mathfrak h = A_4$ and $A_2\oplus A_1$, the rank inequality 
is saturated and there is one $r_i$ even, so \eqref{ex_Drodd_2} 
still gives the no-anomaly condition for $k$.
$D_5$ admits also $D_3\oplus D_2\cong A_3 \oplus 2 A_1$, 
$A_1\oplus D_3\cong A_3 \oplus A_1$, $A_2\oplus D_2\cong 
A_2 \oplus 2 A_1$, $2D_2\cong 4A_1$ and $A_1\oplus D_2\cong 3 A_1$,
see \cite{Lorente} or the method described above,
where only the left hand sides respect the inequality for ranks
and should be used to extract the no-anomaly conditions.
For $A_2\oplus D_2$, $2D_2$ and $A_1\oplus D_2$ either the inequality
for ranks is saturated and there is an even $r_i$ or the inequality
for ranks is strict, hence there are no anomalies for levels
satisfying \eqref{ex_Drodd_2}. Finally, for $D_3\oplus D_2$ and $A_1\oplus D_3$ 
the rank inequality is saturated by there is no even $r_i$ and the
gauge anomalies persist for 
$Z \cong \mathbb Z_4$ if $k \in 2\mathbb Z \setminus 4 \mathbb Z$ 
and for $Z \cong \mathbb Z_2$ if $k$ odd.

\paragraph{Case of $r$ even.}

Taking $M$ and $\tilde{M}$ as given in \eqref{MMtilde_Dr_even} and 
following the same reasoning as for the case of $r$ odd, we get 
the same conditions: 
\begin{equation}
\frac{\tilde{a}_1(r_i+1)}{2}\in\mathbb Z\,, \qquad i=1,\dots,m\,,
\label{=}
\end{equation}
and, if $r_1 + 1 + \ldots + r_m + 1 + s_1 + 
\ldots s_n < r$, 
\begin{equation}
\frac{\tilde{a}_1}{2}\in\mathbb Z
\label{<}
\end{equation}
Additionally, if there are no $D_{s_i}$ components in $\mathfrak{h}$,
then
\begin{eqnarray}
&&\tilde a_1\frac{r}{2}+\tilde a_2\in2\mathbb Z\hspace{1.6cm}\text{for the }
1^{\rm st}\text{ \,embedding}\,\cr
&&\tilde a_1\big(\frac{r}{2}-1\big)+\tilde a_2\in2\mathbb Z\qquad\text{for 
the }2^{\rm nd}\text{ embedding}\,
\end{eqnarray}
(the last two conditions differ only if all $r_i$ are odd and the
rank inequality is saturated because in the other cases $\tilde a_1$ has
to be even). Examining the quantity 
\eqref{Quantity_Dr_even} which has to be an integer with this information 
in mind and taking into account the admissibility conditions 
for the levels, we deduce 

\begin{prop}
The untwisted  coset models built with Lie algebra $\mathfrak{g} 
=\mathfrak{so}(2r)$, $r$ even, and a regular 
subalgebra  $\mathfrak{h} = A_{r_1}
\oplus \ldots \oplus A_{r_m} \oplus D_{s_1} \oplus 
\ldots \oplus D_{s_n}$ do not
have global gauge anomalies for the following cases 
\begin{itemize}
\item$r_1 + 1 + \ldots + r_m + 1 + s_1 + \ldots s_n = r \text{ with all }
r_i \text{ odd }  \\ \hspace*{0.2cm}k \in \left\lbrace
\begin{array}{l} 
2\mathbb{Z} \text{ for any } Z \\
\mathbb{Z}\ \,\text{ if } r/2 \text { even, no } D_{s_i} \text{ and }Z=Z_1
\ \ \ 
\text{ for the } 1^{\rm st}\text{\, embedding}\\
\mathbb{Z}\ \,\text{ if } r/2 \text { even, no } D_{s_i} \text{ and }Z=Z_{diag}
\text{ for the } 2^{\rm nd}\text{ embedding}
\end{array}\right.$
\item\hspace{-0.3cm} 
$\left.\begin{array}{l}r_1 + 1 + \ldots + r_m + 1 + s_1 + \ldots s_n < r 
\text{ or}\\ r_1 + 1 + \ldots + r_m + 1 + s_1 + \ldots s_n = r
\text{ and some } r_{i}\text{ even}\end{array}
\right\rbrace\\ \hspace*{0.26cm}k \in
\left\lbrace
\begin{array}{l} 
2\mathbb{Z} \text{ if } Z = \tilde{Z},\, Z_1 \text{ or } Z_\text{diag} \\
\mathbb{Z}\ \,\text{ if } Z = Z_2 \\
\mathbb Z\ \,\text{ if } r/2 \text{ even, no } D_{s_i} \text{ and any } Z\\
\end{array}\right.$
\end{itemize}
The other untwisted models with admissible levels are not globally 
gauge invariant.
\end{prop}

\noindent{\bf Remark}\ \ \ In particular, the global gauge anomalies present
if $\mathfrak h=\mathfrak g$ for $Z=Z_2$ and $k$ odd disappear 
for $\mathfrak h=A_{r_1}\oplus \ldots\oplus A_{r_m}
\oplus D_{s_1}\oplus\ldots\oplus D_{s_n}$
if $\,r_1 + 1 + \ldots + r_m + 1 + s_1 + \ldots s_n < r\,$ or if
$\,r_1 + 1 + \ldots + r_m + 1 + s_1 + \ldots s_n = r\,$ with some
$r_i$ even.

\paragraph{Example: $\mathfrak{g}=D_4=\mathfrak{so}(8)$.} Here $r$ 
and $r/2$ are both even, so all levels 
$k \in \mathbb Z$ are admissible for all $Z$ and there are no anomalies 
in the case $\mathfrak h = \mathfrak g$ for $k$ even according
to Proposition \ref{prop_Dreven_h=g}, whereas 
the cases with $k$ odd are anomalous. The possible (proper, nontrivial) 
subalgebras $\mathfrak h$ are: $A_1$, $A_2$, $2A_1$, $A_3$ (the latter 
two with 2 inequivalent embeddings), $D_2$, $D_3$, $2D_2$ and $A_1\oplus D_2$.
Note that the two embeddings of $2A_1$ and that of $D_2$ are 
related by the outer automorphisms of $D_4$ and similarly for 
the two embeddings of $A_3$ and the one of $D_3$. For regular subalgebra 
$\mathfrak h = A_1$ or $A_2$, 
the inequality on ranks is strict and there are no $D_{s_i}$ so there
are no anomalies for $k\in\mathbb Z$ for all $Z$. For $D_2$ or $D_3$,
the rank inequality is still strict and there are no anomalies for $k$
even and all $Z$ and for $k$ odd and $Z=Z_2$. For $A_1\oplus D_2$ or 
$2D_2$, the rank inequality is saturated and there are no anomalies 
for even $k$ and any $Z$. Finally, for $2A_1$ or $A_3$ the rank inequality is 
saturated and there are no $D_{s_i}$ so there are no anomalies for
$k$ even and any $Z$ and for $k$ odd and $Z=Z_1$ for the 1$^{\rm st}$ embedding
and $Z=Z_{diag}$ for the 2$^{\rm nd}$ one. 
\vskip 0.4cm

Recall from Sec.\,\ref{sec:D_reven_tw} that the twisted coset
models for $\mathfrak g=\mathfrak{so}(2r)=\mathfrak h$ 
with $r>4$ even have gauge 
anomalies for  $Z=\tilde Z$ (- theory) if $k$ is even and for 
$Z=\tilde Z,\,Z_1$ or $Z_{diag}$ if $k$ is odd for $r/2$ even. 
These are the cases where the no-anomaly condition (\ref{Dreven_tw_c}) 
may be violated. The restriction $e^{2\ii\pi\tilde M}\in\tilde H$
for $\mathfrak h=A_{r_1}\oplus \ldots\oplus A_{r_m}
\oplus D_{s_1}\oplus\ldots\oplus D_{s_n}$
if $\,r_1 + 1 + \ldots + r_m + 1 + s_1 + \ldots s_n < r\,$ or if
$\,r_1 + 1 + \ldots + r_m + 1 + s_1 + \ldots s_n = r\,$ with some
$r_i$ even imposes the condition $\tilde a_1\in 2\mathbb Z$ removing 
the anomalies in the case $Z=\tilde Z$ (- theory) for $k$ even and, if,
additionally, there are no $D_{s_i}$ components in $\mathfrak{h}$, 
also for $Z\not=Z_2$ and $k$ odd. If there are no $D_{s_i}$ and
$r_1+1+\ldots+r_m+1=r$ with all $r_i$ odd then for $k$ odd ($r/2$ even)
the anomalies for $Z=\tilde Z$ are removed for the $+$ theory in the
case of the 1$^{\rm st}$ embedding and for the $-$ theory in the case of
the 2$^{\rm nd}$ embedding, and for $Z=Z_1,\ Z_{diag}$ in the case 
of both embeddings. We obtain this way 

\begin{prop}
The twisted  coset models built with Lie algebra $\mathfrak{g} 
=\mathfrak{so}(2r)$, $r>4$ even, and a regular subalgebra  
$\mathfrak{h} = A_{r_1}\oplus \ldots \oplus A_{r_m} \oplus D_{s_1} 
\oplus \ldots \oplus D_{s_n}$ do not have global gauge anomalies for 
the following cases 
\begin{itemize}
\item$r_1 + 1 + \ldots + r_m + 1 + s_1 + \ldots s_n = r \text{ with all }
r_i \text{ odd }  \\ \hspace*{0.2cm}k \in \left\lbrace
\begin{array}{l} 
2\mathbb{Z} \text{ if } Z=\tilde Z \text{ (+ theory) or } Z=Z_1,\,Z_{diag} 
\\
\mathbb{Z}\ \,\text{ if } Z=Z_2\\
\mathbb{Z}\ \,\text{ if } r/2 \text { even, no } D_{s_i} \text{ and }Z=\tilde Z
\text{ (+ theory) for the } 1^{\rm st}\text{\, embedding}\\
\mathbb{Z}\ \,\text{ if } r/2 \text { even, no } D_{s_i} \text{ and }Z=\tilde Z
\text{ \,(- theory) \hspace{0.07cm}for the } 2^{\rm nd}\text{ embedding}\\
\mathbb{Z}\ \,\text{ if } r/2 \text{ even, no } D_{s_i} \text{ and }
Z=Z_1,\,Z_{diag}
\end{array}\right.$
\item\hspace{-0.3cm} 
$\left.\begin{array}{l}r_1 + 1 + \ldots + r_m + 1 + s_1 + \ldots s_n < r 
\text{ or}\\ r_1 + 1 + \ldots + r_m + 1 + s_1 + \ldots s_n = r
\text{ and some } r_{i}\text{ even}\end{array}
\right\rbrace\\ \hspace*{0.26cm}k \in
\left\lbrace
\begin{array}{l} 
2\mathbb{Z} \text{ if } Z = \tilde{Z},\, Z_1 \text{ or } Z_\text{diag} \\
\mathbb{Z}\ \,\text{ if } Z = Z_2 \\
\mathbb Z\ \,\text{ if } r/2 \text{ even, no } D_{s_i} \text{ and } 
Z=\tilde Z,\,Z_1,\,Z_{diag}
\end{array}\right.$
\end{itemize}
The other twisted models with admissible levels are not globally 
gauge invariant.
\end{prop}

The above results also hold for the coset model with 
$\mathfrak g=\mathfrak{so}(8)$ with twist $\omega_1$,
see (\ref{omegas}). Hence, for $\mathfrak h=A_1$ or $A_2$
there are no gauge anomalies. For $D_2$ or $D_3$ there are
no anomalies if $k$ is even for any $Z$ and if $k$ is odd
for $Z=Z_2$. For $A_1\oplus D_2$ or $2D_2$ there are no anomalies
if $k$ is even for $Z=\tilde Z$ (+ theory) or $Z=Z_1,\,Z_2,\,Z_{diag}$
or if $k$ is odd and $Z=Z_2$. Finally, for $2A_1$ or $A_3$ there are no
anomalies for $Z=\tilde Z$ (+ theory for the 1$^{\rm st}$ embedding, 
- theory for the 2$^{\rm nd}$ one) and for $Z=Z_1,Z_2,Z_{diag}$. In accordance
with the discussion of Sec.\,\ref{sec:D_reven_tw}, we may obtain
the result for twist $\omega_2$ from the one for $\omega_1$
by applying the permutation $Z\rightarrow\omega_4(Z)$ induced 
by the outer automorphism $\omega_4$ on the cyclic subgroups 
of $\tilde Z$, see Eqs.\,(\ref{permutZ}), and on the one
$\mathfrak h\rightarrow\omega_4(\mathfrak h)$  
on subalgebras (modulo inner automorphisms) induced by the
action (\ref{permutalpha}) of $\omega_4$ on simple roots:
\begin{equation}
\begin{array}{ll}\label{permuth}
&\hspace*{-0.6cm}\omega_4(A_1)=A_1,\ \,\omega_4(A_2)=A_2,\ \,
\omega_4((2A_1)^{(1)})
=(2A_1)^{(2)},\ \,\omega_4((2A_1)^{(2)})=D_2,\ \,
\omega_4(A_3^{(1)})=A_3^{(2)},\\
&\hspace*{-0.6cm}\omega_4(A_3^{(2)})=D_3,\ \,\omega_4(D_2)=(2A_1)^{(1)},\ \,
\omega_4(D_3)=A_3^{(1)},\ \,\omega_4(2D_2)=2D_2,\ \,
\omega_4(A_1\oplus D_2)=A_1\oplus D_2.
\end{array}
\end{equation}
where the superscript $(i),\ i=1,2$, labels the independent embeddings.
Similarly, the result for twist $\omega_3$ from the one for $\omega_1$
by applying the inverse permutations $Z\rightarrow\omega_4^{-1}(Z)$
and $\mathfrak h\rightarrow\omega_4^{-1}(\mathfrak h)$.
\,For twists $\omega_4,\omega_4^{-1}$, the the remaining
gauge anomalies are lifted if $\mathfrak h=A_1$ or $A_2$ 
imposing the restrictions $\tilde a_1,\tilde a_2\in 2\mathbb Z$
resulting in 
\begin{prop}
The twisted coset models built with Lie algebra $\mathfrak{g} =
\mathfrak{so}(8)$ with twist $\omega_4$ have global gauge anomalies 
for regular subalgebras 
$\mathfrak{h}=2A_1$,\,$A_3$,\,$D_2$,\,$D_3$,\,$2D_2$,\,$A_1\oplus D_2$ 
and $Z=\tilde Z$ (- theory). The other cases of coset models with 
Lie algebra $\mathfrak{so}(8)$ and twist $\omega_4$ are without anomalies.
\end{prop}

\noindent Similarly

\begin{prop}
The twisted coset models built with Lie algebra $\mathfrak{g} =
\mathfrak{so}(8)$ with twist $\omega_4^{-1}$ have global gauge anomalies 
for regular subalgebras $\mathfrak{h}=2A_1$,\,$A_3$,\,$D_2$,\,$D_3$,\,$2D_2$
and $A_1\oplus D_2$ and $Z=\tilde Z$ (($-$)$^k$ theory). The other cases 
of coset models with Lie algebra $\mathfrak{so}(8)$ and twist $\omega_4^{-1}$ 
are without anomalies.
\end{prop}

\subsection{Regular semisimple subalgebras of $\mathfrak e_{6}$}

In this case with fixed rank $r=6$, one can establish a complete list 
of regular semisimple subalgebras, up to conjugation, with an embedding, 
however, that is not explicit \cite{Dynkin,Lorente}. We shall only need 
the embedding of simple roots
in the ambient algebra which is enough to reconstruct the full embedding 
using the Serre construction. The element $M$ and $\tilde{M}$ will be
described employing the explicit
realization of the coweight and coroot lattices of $\mathfrak{e}_6$,
\begin{equation}\label{Coweight_Lattice_e6}
 P^\vee(\mathfrak{e}_6) = \left\lbrace \left( \dfrac{a}{6} + q_1, \ldots,
\dfrac{a}{6} + q_6, \dfrac{b}{\sqrt{2}} \right) \left| 
\begin{array}{l}a,b, q_1,
\ldots, q_6 \in \mathbb{Z}\\ a + q_1 + \ldots + q_6 = 0\\ a + b \in 2\mathbb{Z}
\end{array} \right. \right\rbrace
\end{equation}
and the coroot lattice $Q^\vee(\mathfrak{e}_6)$ is defined the same way but
adding the condition $a \in 3 \mathbb{Z}$. We shall consider only the
untwisted coset models because the twisted ones are non-anomalous,
see Proposition \ref{prop_e6_h=g_tw}.
Taking $M$ and $\tilde{M}$ in
$P^\vee(\mathfrak{e}_6)$ with the corresponding coefficients, the quantity
\eqref{Quantity_e6} becomes
\begin{equation}\label{Quantity_e6_bis}
 k~\text{tr}(M\tilde{M}) = k \dfrac{a \tilde{a}}{3} + m, \qquad \text{ with }
m \in \mathbb{Z}
\end{equation}
Now, specifying a subalgebra $\mathfrak{h}\subset \mathfrak e_{6}$ 
and requiring that $e^{2\ii \pi
\tilde{M}} \in \tilde{Z}\cap\tilde{H}$, two possibilities arise: if one can
show that $\tilde{a} \in 3\mathbb{Z}$ then the previous quantity is an integer
for
every $k \in \mathbb{Z}$ and all the corresponding coset models are globally
gauge invariant. Otherwise, if there exist an element $\tilde{M}$ such that
$\tilde{a} \notin 3\mathbb{Z}$, then we have to require $k \in 3\mathbb{Z}$ to
have a globally gauge invariant coset model, and the other coset models 
are anomalous. Before examining the anomaly problem for every regular
subalgebra of $\mathfrak{e}_6$, one can make four remarks:

\begin{itemize}
\item if there are no anomalies for a given subalgebra $\mathfrak{h}$ of
$\mathfrak{e}_6$ ($\tilde{a} \in 3\mathbb{Z}$), then the regular subalgebras 
that are smaller (and will be obtained from the Dynkin diagram of 
$\mathfrak{h}$ by the procedure described above) lead also
to the condition $\tilde{a} \in 3\mathbb{Z}$, inheriting 
it from $\mathfrak{h}$. In other words, the regular subalgebra with 
no anomalies protects the cases of its regular subalgebras. Consequently, 
we will look only at the cases where the anomalies are present and treat 
the problem by decreasing rank.

\item Among the regular subalgebras generated by the algorithm described
at the beginning of Sec.\,\ref{sec:4},
many can still be mapped into each other by the conjugations 
that normalize $\,\mathfrak t_{\mathfrak e_6}\,$ (and induce on it Weyl group 
transformations) and, as a result, they lead to the same condition 
for the absence of anomalies. We may then consider only one regular 
subalgebra in each class of subalgebras related by Weyl group transformations.
In particular, there are Weyl group transformations that permute the simple 
roots $\,\alpha_i\,$ and $\,\delta=-\phi\,$ according to the symmetries of the 
extended Dynkin diagrams (see, e.g., Appendix B of \cite{Gawedzki1})
and they permit to restrict the count of regular subalgebras. 

\item The subalgebras related by the outer automorphism of $\mathfrak{e}_6$ 
lead to the same no-anomaly condition, see the remark at the end of
Sec.\,\ref{sec:2}.

\item Since $e^{2\ii \pi\tilde{M}} \in\tilde Z\cap \tilde{H}$ if and only if
$\tilde M\in P^\vee(\mathfrak g)$ and $\tilde{M}+q^\vee\in\ii
\mathfrak{t}_{\mathfrak h}\subset\ii\mathfrak{t}_\mathfrak{g}$ for some
$q^\vee\in Q^\vee(\mathfrak{e}_6)$, it is enough to check the no-anomaly
condition (\ref{coset_anomaly}) only for $\tilde{M}\in P^\vee(\mathfrak g)$
perpendicular to the orthogonal complement $\ii\mathfrak{t}^\bot_{\mathfrak h}$
of $\ii \mathfrak{t}_{\mathfrak h}$ in $\ii\mathfrak{t}_{\mathfrak g}$.
%
\end{itemize}

We now consider the regular semisimple subalgebras, beginning 
by those of rank 6 and then decreasing the rank. Subspace 
$\ii\mathfrak{t}_\mathfrak{h}^\bot$  (which is small for high
ranks) is computed for each subalgebra and we look at the consequences 
of the condition $\tilde M\perp \ii\mathfrak{t}_\mathfrak{h}^\bot$ 
on $\tilde{M}$. Upon using the protection 
property and the Weyl transformations described above, as well as
the outer automorphism of $\mathfrak{e}_6$, only a few cases have 
to be treated. The explicit computation is given in Table 
\ref{Computation_e6_regular}. The subalgebras of rank 6 are not represented 
because we have $\mathfrak{t}_\mathfrak{h}^\bot = \emptyset$, so there is 
no supplementary condition for $\tilde{M}$ and there are always anomalies if 
$k \notin 3\mathbb{Z}$. Only subalgebras of rank 5 and 4 have 
potential anomalies, the ones of lower ranks being protected by 
a possible inclusion into non-anomalous subalgebras. 
\begin{table}[htb]
\centering
\begin{tabular}{|c|c|c|c|}
\hline
   $\mathfrak{h}$ & simple roots of $\mathfrak{h}$ &
basis of $\ii\mathfrak{t}_{\mathfrak{h}}^\bot$ & $\tilde{M}$ \\
\hline \hline
 $D_5$ &  $\alpha_1, \alpha_2, \alpha_3, \alpha_4,
\alpha_6$ & $\left(1,1,1,1,1,\text{-}5, 3\sqrt{2} \right)$ &
$\tilde{a} \in 3\mathbb{Z} $ \\
\hline
 $A_3 \oplus 2A_1$ &  $\alpha_1, \alpha_2, \alpha_3
\oplus \delta \oplus \alpha_5$ & $\left(1,1,1,1,\text{-}2,\text{-}2, 0
\right)$ & $\tilde{a} \in 3\mathbb{Z} $ \\
\hline 
  $A_4 \oplus A_1$ &  $\alpha_1, \alpha_2, \alpha_3,
\alpha_4 \oplus \delta$ & $\left(1,1,1,1,1,\text{-}5, 0 \right)$ &
$\tilde{a} \in 3\mathbb{Z} $\\
\hline
  $A_5$ &  $\alpha_1, \alpha_2, \alpha_3, \alpha_4, \alpha_5$
& $\left(0,0,0,0,0,0, 1 \right)$ & $\tilde{a}\in 2\mathbb Z$ \\
\hline 
 $2A_2\oplus A_1$ &  $\alpha_1, \alpha_2 \oplus
\alpha_4,\alpha_5 \oplus \alpha_6$ & $\left(1,1,1,-1,-1,-1,3\sqrt{2}
\right)$ & $\tilde{a}\in\mathbb Z$ \\
\hline \hline
 $2A_2$ & $\alpha_1, \alpha_2 \oplus \alpha_4,
\alpha_5$ & $\left(1,1,1,\text{-}1,\text{-}1,\text{-}1, 0 \right)$ &
$\tilde{a}\in2\mathbb Z$\\
 & & $\left(0,0,0,0,0,0,1 \right)$ & \\
\hline
 \end{tabular}
\caption{$\ii\mathfrak{t}_\mathfrak{h}^\bot$ for the regular
subalgebras of $\mathfrak{e}_6$ of rank $5$ and $4$ and consequences 
for $\tilde a$; the simple roots $\alpha_i$ of $\mathfrak{e}_6$ and 
its lowest root $\delta$ are used to generate the regular subalgebras
\cite{Lorente}.}\label{Computation_e6_regular}
\end{table}

\noindent We are thus able to state

\begin{prop}
The untwisted coset models built with Lie algebra $\mathfrak{g} 
= \mathfrak{e}_6$ 
and any regular subalgebra $\mathfrak{h}$ do not have global gauge anomalies 
for every $k\in\mathbb{Z}$, except for the cases 
$\mathfrak{h} = \mathfrak{e}_6, A_5 \oplus A_1, 3A_2$, 
of rank 6, $A_5, 2A_2\oplus A_1$,
of rank 5, and $2A_2$ of rank 4, where the only globally gauge
invariant models are those with $k \in 3 \mathbb{Z}$. 
\end{prop}

\section{R-subalgebras and S-subalgebras}
\label{sec:5}

The regular subalgebras are not the only possible Lie subalgebras for a given
ambient Lie algebra. We can use them, however, to classify all the remaining 
ones. Let $\mathfrak{h}$ be a semisimple subalgebra of $\mathfrak{g}$. 
Let $\mathcal{R}(\mathfrak{h})$ be a minimal regular
subalgebra of $\mathfrak{g}$ containing $\mathfrak h$ (up to conjugation). 
If $\mathcal{R}(\mathfrak{h}) =\mathfrak{g}$, then 
$\mathfrak{h}$ is called an S-subalgebra. Otherwise, 
it is called an R-subalgebra. For the exceptional simple algebras, the 
classification of R- and S-subalgebras has been achieved by Dynkin 
in \cite{Dynkin}. 
The case of other simple algebras was discussed in \cite{Dynk2} with
less explicit results. In this section, we first treat completely the 
case of non-regular subalgebras of the exceptional Lie algebra 
$\mathfrak{g} = \mathfrak{e}_6$ which may have anomalies and then we consider 
some examples of non-regular subalgebras of classical Lie algebras.

\paragraph{Dynkin index.} Consider a simple Lie subalgebra $\mathfrak
h \subset \mathfrak g$ of a semisimple Lie algebra $\mathfrak g$ 
and the corresponding embedding $\iota$. \,The relation 
\begin{equation}
  {\rm tr}_{\mathfrak{g}}(\iota(X))^2\,=\,j\,{\rm tr}_{\mathfrak{h}}X^2 \qquad {\rm for}\quad X \in \mathfrak h\,
\end{equation}
\noindent where the invariant quadratic forms $\,{\rm tr}_{\mathfrak g}\,$ 
and $\,{\rm tr}_{\mathfrak h}\,$ have the normalizations described in
the beginning of Sec.\,\ref{sec:2}, 
defines the scalar factor $\,j\,$ (independent of $X$), 
 called Dynkin index, which
is always an integer \cite{Dynkin}. Moreover, $\,j\,$ is invariant under
composition of $\,\iota\,$ with inner (and outer) automorphisms of 
$\mathfrak g$, so that it depends on the class of equivalent 
embeddings. 

\subsection{Simple nonregular subalgebras of $\mathfrak{e}_6$}
\label{sec:SimpleCase}

\paragraph{Subalgebras of rank 1}

According to Dynkin, the subalgebra $\mathfrak{h} = A_1$ can be embedded in
several different ways in $\mathfrak{e}_6$, as regular, R- and S-subalgebra and
the embedding $\iota$ is fully characterized by the embedding 
of the simple coroot $\alpha^\vee$ of $A_1$. Recall the compatibility 
condition for $\tilde{M}$ in the anomaly problem

\begin{equation}
 e^{2\ii \pi \tilde{M}} \in \tilde{H} \cap \tilde{Z} \subseteq
\mathcal{Z}(\tilde{H})\,,
\end{equation}
where  $\mathcal{Z}(\tilde{H})=\{1,e^{2\ii\pi\iota(\lambda^\vee)}\}$ 
with $\lambda^\vee =\frac{1}{2} \alpha^\vee$ is the center of $\tilde H$
which is either trivial (if $1=e^{2\ii\pi\iota(\lambda^\vee)}$ and 
$\tilde H\cong SO(3)$) or is isomorphic to $\mathbb Z_2$
(if $1\not=e^{2\ii\pi\iota(\lambda^\vee)}$ and $\tilde H\cong SU(2)$). 
Looking at the embedding of $\lambda^\vee$ in $\mathfrak{e}_6$, three
possibilities can occur

\begin{enumerate}
 \item If $\iota(\lambda^\vee) \notin P^\vee(\mathfrak{e}_6)$ 
then $\tilde{Z}
\cap \tilde{H} = \lbrace 1 \rbrace$ and $\tilde{M}$ is a coroot of
$\mathfrak{e}_6$, so the quantity \eqref{Quantity_e6} is always 
an integer and there are no anomalies for this model.
\item If $\iota(\lambda^\vee) \in Q^\vee(\mathfrak{e}_6)$ 
then $\tilde{M}$ is still
only a coroot of $\mathfrak{e}_6$, and there are no anomalies too.
\item If $\,\iota(\lambda^\vee) \in P^\vee(\mathfrak{e}_6)\setminus 
Q^\vee(\mathfrak{e}_6)$ then anomalies are
possible and we have to check that the quantity \eqref{Quantity_e6}
is an integer for $\tilde{M} = \iota(\lambda^\vee)$ looking at the
corresponding value for $\tilde{a}$, see Eq.\,(\ref{Quantity_e6_bis}).
\end{enumerate}
The explicit embeddings are given in \cite{Dynkin} (Chapter III, Table 18), and
the computation of the intersection with the roots of $\mathfrak{e}_6$ is done
in Table \ref{Computation_e6_rank1} for each subalgebra of rank 1: the
possibility 3 never occurs, so there are no anomalies for 
the corresponding coset models for any $k \in \mathbb{Z}$.

\begin{table}[htb]
\centering 
\begin{tabular}{|c|c|c|c|}
\hline
$\mathcal{R}(\mathfrak{h})$ & Index & $\iota(\lambda^\vee)$ & Compatibility\\
\hline \hline
$A_1$ & 1 & $ \left(0,0,0,0,0,0,\frac{1}{\sqrt{2}}\right)$ & $\notin P^\vee(\mathfrak{e}_6)$\\
\hline \hline
$2 A_1$ & 2 &$ \left( \tfrac{1}{2},0,0,0,0,\tfrac{\text{-}1}{2},\tfrac{1}{\sqrt{2}}
\right)$ & $\notin P^\vee(\mathfrak{e}_6)$\\
\hline
$3 A_1$ & 3 & $ \left(
\tfrac{1}{4},\tfrac{1}{4},\tfrac{1}{4},\tfrac{\text{-}1}{4},\tfrac{\text{-}1}{4}
,\tfrac{\text{-}1}{4},\tfrac{3}{2\sqrt{2}} \right)$ & $\notin
P^\vee(\mathfrak{e}_6)$\\
\hline
$ A_2$ & 4 &$ \left( 0,0,0,0,0,0,\sqrt{2} \right)$ & $\in Q^\vee(\mathfrak{e}_6)$\\
\hline
$ A_2 \oplus A_1$ & 5 & $ \left(
\tfrac{1}{2},0,0,0,0,\tfrac{\text{-}1}{2},\sqrt{2} \right)$ & $\notin
P^\vee(\mathfrak{e}_6)$\\
\hline
$ A_2 \oplus 2A_1$ & 6 & $ \left(
\tfrac{1}{2},\tfrac{1}{2},0,0,\tfrac{\text{-}1}{2},\tfrac{\text{-}1}{2},\sqrt{2}
\right)$ & $\notin P^\vee(\mathfrak{e}_6)$\\
\hline
$ 2A_2$ & 8 &$ \left(1,0,0,0,0,\text{-}1,\sqrt{2}\right)$ & $\in Q^\vee(\mathfrak{e}_6)$\\
\hline
$2 A_2 \oplus A_1$ & 9 & $ \left(
\tfrac{3}{4},\tfrac{1}{4},\tfrac{1}{4},\tfrac{\text{-}1}{4},\tfrac{\text{-}1}{4}
,\tfrac{\text{-}3}{4},\tfrac{5}{2\sqrt{2}} \right)$ & $\notin
P^\vee(\mathfrak{e}_6)$\\
\hline
$ A_3 $ & 10 & $ \left( \tfrac{1}{2},0,0,0,0,\tfrac{\text{-}1}{2},\tfrac{3}{\sqrt{2}}
\right)$ & $\notin P^\vee(\mathfrak{e}_6)$\\
\hline
$ A_3 \oplus A_1$ & 11 & $ \left(
\tfrac{1}{2},\tfrac{1}{2},0,0,\tfrac{\text{-}1}{2},\tfrac{\text{-}1}{2},\tfrac{3
}{\sqrt{2}} \right)$ & $\notin P^\vee(\mathfrak{e}_6)$\\
\hline
$ A_3 \oplus 2A_1$ & 12 & $ \left(
\tfrac{1}{2},\tfrac{1}{2},\tfrac{1}{2},\tfrac{\text{-}1}{2},\tfrac{\text{-}1}{2}
,\tfrac{\text{-}1}{2},\tfrac{3}{\sqrt{2}} \right)$ & $\in Q^\vee(\mathfrak{e}_6)$\\
\hline
$ A_4$ & 20 & $ \left( 1,0,0,0,0,\text{-}1,2\sqrt{2} \right)$ & $\in Q^\vee(\mathfrak{e}_6)$\\
\hline
$ A_4 \oplus A_1$ & 21 & $
\left(1,\tfrac{1}{2},0,0,\tfrac{\text{-}1}{2},\text{-}1,2\sqrt{2}
\right)$ & $\notin P^\vee(\mathfrak{e}_6)$\\
\hline
$ D_4$ & 28 & $ \left(
\tfrac{1}{2},\tfrac{1}{2},\tfrac{1}{2},\tfrac{\text{-}1}{2},\tfrac{\text{-}1}{2}
,\tfrac{\text{-}1}{2},\tfrac{5}{\sqrt{2}} \right)$ & $\in Q^\vee(\mathfrak{e}_6)$\\
\hline
$ D_5 (a_1)$ & 30 & $
\left(1,\tfrac{1}{2},0,0,\tfrac{\text{-}1}{2},\text{-}1,\tfrac{5}{\sqrt{2}}
\right)$ & $\notin P^\vee(\mathfrak{e}_6)$\\
\hline
$ A_5$ & 35 & $ \left(
\tfrac{3}{2},\tfrac{1}{2},0,0,\tfrac{\text{-}1}{2},\tfrac{\text{-}3}{2},\tfrac{5
}{\sqrt{2}}\right)$ & $\notin P^\vee(\mathfrak{e}_6)$\\
\hline
$ A_5 \oplus A_1$ & 36 & $ \left(
\tfrac{3}{2},\tfrac{1}{2},\tfrac{1}{2},\tfrac{\text{-}1}{2},\tfrac{\text{-}1}{2}
,\tfrac{\text{-}3}{2},\tfrac{5}{\sqrt{2}}\right)$ & $\in Q^\vee(\mathfrak{e}_6)$\\
\hline
$D_5 $ & 60 & $ \left(
\tfrac{3}{2},\tfrac{1}{2},\tfrac{1}{2},\tfrac{\text{-}1}{2},\tfrac{\text{-}1}{2}
,\tfrac{\text{-}3}{2},\tfrac{7}{\sqrt{2}}\right)$ & $\in Q^\vee(\mathfrak{e}_6)$\\
\hline \hline
$ \mathfrak{e}_6 (a_1)$ & 84 &$
\left(2,1,0,0,\text{-}1,\text{-}2,4\sqrt{2}\right)$ & $\in Q^\vee(\mathfrak{e}_6)$\\
\hline
$\mathfrak{e}_6 $ & 156 & $ \left(
\tfrac{5}{2},\tfrac{3}{2},\tfrac{1}{2},\tfrac{\text{-}1}{2},\tfrac{\text{-}3}{2}
,\tfrac{\text{-}5}{2},\tfrac{11}{\sqrt{2}}\right)$ & $\in Q^\vee(\mathfrak{e}_6)$\\
\hline 
  \end{tabular}
\caption{The embedding of element $\lambda^\vee$ for rank 1
subalgebras and its intersection with the lattices of
$\mathfrak{e}_6$.}
\label{Computation_e6_rank1}
\end{table}

\paragraph{Simple S-subalgebras of rank > 1}

Following \cite{Dynkin} (Chapter IV, Table 24), there exist four S-subalgebras
of $\mathfrak{e}_6$ of rank $>1$: $\mathfrak{h} = A_2, \mathfrak{g}_2, C_4$ 
and $\mathfrak{f}_4$. For the cases $\mathfrak{g}_2$ and $\mathfrak{f}_4$, 
the center
of the corresponding group is $\mathcal{Z}(\tilde{H}) \cong \lbrace 1 \rbrace$.
Then $\tilde{M}$ can be only a coroot of $\mathfrak{e}_6$ and the quantity
\eqref{Quantity_e6} is always an integer. For the two remaining cases, the
explicit
embedding is still given in \cite{Dynkin}, and the strategy is the same as for
rank one: we look how the generating element $\iota(\lambda^\vee)$ of
$\mathcal{Z}(\tilde{H})$ intersects with the lattices of $\mathfrak{e}_6$ and
check which possibility occurs among those listed in the case of rank one
(except that we would also have to check that for the low multiples 
of $\lambda^\vee$ if $\iota(\lambda^\vee)$ were
not in $Q^\vee(\mathfrak{g})$). The results are described in Table 
\ref{Computation_e6_S} from which we infer that there are no gauge anomalies 
for all simple S-subalgebras of $\mathfrak e_6$.

\begin{table}[htb]
 \centering
\begin{tabular}{|c|c|c|c|c|}
  \hline
$\mathfrak{h}$ & $\mathcal{R}(\mathfrak{h})$ & Index & $\iota(\lambda^\vee)$ &
Compatibility \\
\hline \hline
$A_2$ & $\mathfrak{e}_6$ & 9 & $ \left(\tfrac{1}{2},\tfrac{1}{2},\tfrac{1}{2},
\tfrac{\text{-}1}{2},\tfrac{\text{-}1}{2},\tfrac{\text{-}1}{2},
\tfrac{3}{\sqrt{2}}\right)$ & $\in Q^\vee(\mathfrak{e}_6)$\\
\hline
$C_4$ & $\mathfrak{e}_6$ & 1 & $ \left(
\tfrac{1}{2},\tfrac{1}{2},\tfrac{1}{2},\tfrac{\text{-}1}{2},\tfrac{\text{-}1}{2}
,\tfrac{\text{-}1}{2},\tfrac{1}{\sqrt{2}}\right)$ &
$\in Q^\vee(\mathfrak{e}_6)$\\
\hline
\end{tabular}
\caption{The embedding of element $\lambda^\vee$ for simple S-subalgebras of
$\mathfrak{e}_6$ and its intersection with the lattices.}
\label{Computation_e6_S}
\end{table}

\paragraph{Simple R-subalgebras of rank > 1}\label{SimpleR}

We only need to look at the R-subalgebras $\mathfrak{h}$ with potential 
anomalies. Indeed, the subalgebra $\mathcal{R}(\mathfrak{h})$ is 
regular, so has been already treated. If $\mathcal{R}(\mathfrak{h})$ 
corresponds to a model without anomalies, then it protects also 
the R-subalgebra $\mathfrak{h}$ included in it and there will be no anomalies 
for the model built with $\mathfrak{h}$. The list of the R-subalgebras of 
$\mathfrak{e}_6$ is
given in \cite{Dynkin} (Chapter IV, Table 25), but without explicit embedding.
There remain five cases with potential anomalies: $\mathfrak{h} = A_2$, with
$\mathcal{R}(\mathfrak{h}) = A_5$,\,$2A_2$,\,$3A_2$, and  $\mathfrak{h} = A_3$ 
or $C_3$ with $\mathcal{R}(\mathfrak{h}) = A_5$. If
$\mathcal{R}(\mathfrak{h})$ is simple, then the embedding of $\mathfrak{h}$ in
$\mathcal{R}(\mathfrak{h})$ is given in \cite{Lorente} (Table XIII), 
considering
$\mathfrak{h}$ as an S-subalgebra of $\mathcal{R}(\mathfrak{h})$.

\begin{table}[htb]
\centering
\begin{tabular}{|c|c|c|c|c|c|}
  \hline
$\mathfrak{h}$ & $\mathcal{R}(\mathfrak{h})$ & Index & $\iota(\lambda^\vee)$ &
Compatibility & $\tilde{a}$ \\
\hline \hline
$A_2$ & $2A_2(\iota_1)$ & 2 & $ \left(
\tfrac{1}{3},\tfrac{1}{3},\tfrac{\text{-}2}{3},\tfrac{2}{3},\tfrac{\text{-}1}{3}
,\tfrac{\text{-}1}{3},0\right)$ &
$\notin P^\vee(\mathfrak{e}_6)$ &\\
\hline
$A_2$  & $2A_2 (\iota_2)$ & 2 & $ \left(
\tfrac{1}{3},\tfrac{1}{3},\tfrac{\text{-}2}{3},\tfrac{1}{3},\tfrac{1}{3},\tfrac{
\text{-}2}{3},0\right)$ &
$\in P^\vee(\mathfrak{e}_6)\setminus Q^\vee(\mathfrak{e}_6)$ & 2\\
\hline
$A_2$ & $3A_2(\iota_1)$ & 3 & $ \left(
0,0,-1,1,0,0,0\right)$ &
$\in Q^\vee(\mathfrak{e}_6)$ &\\
\hline
$A_2$ & $3A_2(\iota_2)$ & 3 & $ \left(0,0,-1,\tfrac{2}{3},\tfrac{2}{3}
,\tfrac{\text{-}1}{3},0\right)$ &
$\notin P^\vee(\mathfrak{e}_6)$ & \\
\hline
$A_2$ & $A_5$ & 5& $ \left(
\tfrac{2}{3},\tfrac{2}{3},\tfrac{\text{-}1}{3},\tfrac{2}{3},\tfrac{\text
{-}1}{3},\tfrac{\text{-}4}{3},0\right)$ &
$\in P^\vee(\mathfrak{e}_6)\setminus Q^\vee(\mathfrak{e}_6)$ & 4\\
\hline
$A_3$ & $A_5$ & 2 & $ \left(
\tfrac{1}{2},\tfrac{1}{2},\tfrac{\text{-}1}{2},\tfrac{1}{2},\tfrac{\text{-}1}{2}
,\tfrac{\text{-}1}{2},0\right)$ & $\notin P^\vee(\mathfrak{e}_6)$ & \\
\hline
$C_3$ & $A_5$ & 1 & $ \left(
\tfrac{1}{2},\tfrac{1}{2},\tfrac{1}{2},\tfrac{\text{-}1}{2},\tfrac{\text{-}1}{2},\tfrac{\text{-}1}{2},0\right)$ &
$\notin P^\vee(\mathfrak{e}_6)$ & \\
\hline
\end{tabular}
\caption{The embedding of element $\lambda^\vee$ for simple R-subalgebras of
$\mathfrak{e}_6$ and its intersection with the lattices. In case of potential
anomalies, the explicit value of $\tilde{a}$ that enters quantity
\eqref{Quantity_e6_bis} is given.}
\label{Computation_e6_R}
\end{table}

\noindent If $\mathcal{R}(\mathfrak{h})$ is only semisimple, the problem 
of the embedding is
treated in \cite{Minchenko}, where several inequivalent embeddings of
$\mathfrak{h}$ in $\mathfrak{e}_6$ appear. 
For the $\mathfrak{h} = A_2$ and $\mathcal{R}(\mathfrak{h}) = 3 A_2$, the two
inequivalent embeddings are the following, denoting by $\tilde\alpha_1^\vee$ 
and $\tilde\alpha_2^\vee$ the simple coroots of $A_2$.
\begin{align}
 \iota_1(\tilde\alpha_1^\vee) = \alpha_1^\vee + \alpha_5^\vee 
+ \delta^\vee & \qquad &
\iota_2(\tilde\alpha_1^\vee) = \alpha_1^\vee + \alpha_4^\vee + \delta^\vee\\
 \iota_1(\tilde\alpha_2^\vee) = \alpha_2^\vee 
+ \alpha_4^\vee + \alpha_6^\vee & \qquad &
\iota_2(\tilde\alpha_2^\vee) = \alpha_2^\vee + \alpha_5^\vee + \alpha_6^\vee
\end{align}
where we have exchanged $\alpha^\vee_4$ and $\alpha^\vee_5$. The other possible
exchanges are equivalent to $\iota_1$ or $\iota_2$ \cite{Minchenko}. For
$\mathcal{R}(\mathfrak{h}) = 2A_2$, the two embeddings are given by
similar formulas but with omission of $\alpha^\vee_6$ and $\delta^\vee$. 
Again, in order to find $Z(\tilde H)\cap\tilde Z$, we have to check how 
the generating element $\iota(\lambda^\vee)$ of 
$\mathcal{Z}(\tilde{H})$ intersects 
with the lattices of $\mathfrak{e}_6$. An explicit calculation is done in 
Table \ref{Computation_e6_R}, and this time potential anomalies occur. 
Then, looking at the value of $\tilde{a}$ for $\tilde{M} =  
\iota(\lambda^\vee)$, we deduce an, eventually more restrictive, condition 
on level $k$ required to avoid the anomalies (to exclude the anomalies in 
the case of $A_3\subset A_5$, we also have to observe that $\iota(2\lambda^\vee)
\in Q^\vee(\mathfrak{e}_6)$). 
\eject

\noindent This way, we obtain the general result for 
simple nonregular subalgebras of $\mathfrak{e}_6$

\begin{prop}
The untwisted coset models with $\mathfrak{g} = \mathfrak{e}_6$ and any
simple, nonregular subalgebra $\mathfrak{h}$ do not have global  gauge
anomalies for $k \in \mathbb{Z}$ except for the R-subalgebras 
$\mathfrak{h} = A_2$ with $\mathcal{R}(\mathfrak{h}) = A_5$ and 
$\mathfrak h=A_2$ with $\mathcal{R}(\mathfrak h)=2A_2$ embedded 
via $\iota_2$. For those subalgebras, the global gauge invariance requires 
that $k \in 3\mathbb{Z}$.
\end{prop}

\subsection{Semisimple nonregular subalgebras of $\mathfrak{e}_6$}

Let $\mathfrak h$ be a semisimple subalgebra of $\mathfrak e_6$:
\begin{equation}
 \mathfrak h = \mathop{\oplus}\limits_{i=1}^{n} \mathfrak h_i
\end{equation}
where the $\mathfrak h_i$ are simple, and the corresponding subgroups 
are denoted by $\tilde H_i$. The case $n =1$ has been already treated 
above, so we now deal with $n \geq 2$. First, suppose that one of the 
$\mathfrak h_i$ considered as a simple subalgebra leads to anomalies: 
there exists $\tilde M_i$ such that $e^{2 \ii \pi \tilde{M}_i} \in 
\tilde H_i \cap \tilde Z$ which imposes $k \in 3 \mathbb Z$ to ensure 
that the quantity $(\ref{Quantity_e6})$ is integral. Then, taking 
$\tilde M = \tilde M_i$ but now embedded in $\mathfrak h$,
we shall still have to impose $k \in 3\mathbb Z$ 
to have a globally gauge invariant model with semisimple Lie algebra 
$\mathfrak h$. In other words, semisimple algebras composed of simple 
ideals with at least one leading to anomalies are also anomalous. 
However, the inverse 
is not true: one can have a semisimple subalgebra corresponding to 
an anomalous model with all its simple ideals without any anomaly. For 
example, the model with regular subalgebra $2A_2$ of 
$\mathfrak e_6$ is anomalous for $k \in \mathbb{Z}\setminus 3\mathbb{Z}$ 
whereas the one with $A_2$ (still regular) is globally gauge invariant 
for every $k \in \mathbb{Z}$. Thus we need to check all the cases where 
all the simple ideals correspond to models without anomaly. To do that, 
we need to consider the elements $\sum_{i=1}^n \alpha_i \iota 
(\lambda^\vee_i)$ where $\alpha_i \in \mathbb{Z}$ and $\lambda_i$ are 
the generating elements of the center of the $\tilde H_i$, which have 
all been described above in the simple case (Tables 
\ref{Computation_e6_rank1}, \ref{Computation_e6_S} and 
\ref{Computation_e6_R}), and $\iota: \mathfrak h \rightarrow 
\mathfrak e_6$ is the embedding. Comparing how these elements are 
compatible with the coroot and coweight lattices of $\mathfrak e_6$, 
the anomaly problem is reduced to the three possibilities described in 
the simple case \ref{sec:SimpleCase}. 

\paragraph{S-subalgebras}

In \cite{Dynkin}
(Chapter V, Table 39) one can find all the S-subalgebra of $\mathfrak e_6$ 
and their including relations. It turns out that subalgebra
 $\mathfrak h = \mathfrak g_2 \oplus A_2$ 
(with the explicit embedding given in \cite{Dynkin}, Chapter V, Table 35) 
leads to an anomaly if $k \in \mathbb Z \setminus 3 \mathbb Z$, and that 
the other semisimple nonsimple S-subalgebras of $\mathfrak e_6$ are protected. 

\begin{table}[ht]
\centering
  \begin{tabular}{|c|c|c|c|c|}
  \hline 
$\mathfrak{h}$ & $\mathcal{R}(\mathfrak{h})$ & $\left\lbrace 
\mathcal{R}(\mathfrak{h}_i) \right\rbrace$ & Indices &
 No anomaly for \\
\hline \hline

$A_2 \oplus A_1$ & $A_5 \oplus A_1$ & $A_5, A_1$ & 2,1 &
${k \in 3\mathbb{Z}}$  \\
$A_3 \oplus A_1$ & $A_5 \oplus A_1$ & $A_5, A_1$ & 2,1 &$k \in \mathbb{Z}$  \\
$C_3 \oplus A_1$ & $A_5 \oplus A_1$ & $A_5, A_1$ & 1,1 &$k \in \mathbb{Z}$  \\
$A_1 \oplus A_1$ & $A_5 \oplus A_1$ & $A_5, A_1$ & 35,1&$k \in \mathbb{Z}$  \\
$(A_2 (\iota_1) \oplus A_1) \oplus A_1$ & $A_5 \oplus A_1$ 
& $A_5, A_1$ & 2,3,1 &$k \in
\mathbb{Z}$  \\
$(A_2 (\iota_2) \oplus A_1) \oplus A_1$ & $A_5 \oplus A_1$ & $A_5, A_1$ 
& 2,3,1 &${k \in
3\mathbb{Z}}$  \\
$(2A_1) \oplus A_1$ & $A_5 \oplus A_1$ & $A_5, A_1$ 
& 8,3,1&$k \in \mathbb{Z}$  \\
\hline
$A_1 \oplus (2A_2)$ & $A_2 \oplus (2A_2)$ 
& $A_2, 2A_2 $ & 4,1,1&${k \in
3\mathbb{Z}}$  \\
$A_2 \oplus A_2 (\iota_1) $ & $A_2 \oplus (2A_2)$ 
& $A_2, 2A_2 $ &1,2 &$k \in
\mathbb{Z}$  \\
$A_2 \oplus A_2 (\iota_2)$ & $A_2 \oplus (2A_2)$ 
& $A_2, 2A_2 $ & 1,2&${k \in
3\mathbb{Z}}$  \\
$A_1 \oplus A_2 (\iota_2)$ & $A_2 \oplus (2A_2)$ 
& $A_2, 2A_2 $ & 4,2&${k \in
3\mathbb{Z}}$  \\
\hline
$A_1 \oplus A_1 \oplus A_2$ & $A_2 \oplus A_2 \oplus A_2$ 
& $A_2, A_2, A_2$ & 4,4,1&
$k\in \mathbb{Z}$  \\
\hline
$A_2(\iota_2) \oplus A_1$ & $A_5$ & $A_5 $ 
& 2,3&${k \in 3\mathbb{Z}}$  \\
\hline
$  A_1 \oplus A_2 (\iota_2)$ & $A_1 \oplus (2A_2)
$ & $A_1, 2A_2$ & 1,2&${k \in 3\mathbb{Z}}$  \\
\hline
$A_1 \oplus A_1 \oplus A_2$ & $A_1 \oplus A_2 \oplus A_2$ 
& $A_1, A_2, A_2$  & 1,4,1&$k \in \mathbb{Z}$  \\
\hline
\end{tabular}\small
\caption{Semisimple nonsimple R-subalgebras of
$\mathfrak{e}_6$ with possible anomalies and the conditions
on $k$ required for their absence}
\label{Computation_e6_ssimple}
\end{table}

\paragraph{R-subalgebras}
The end of \cite{Lorente} proposes a method to construct all the
semisimple R-subalgebras: the idea is to take the semisimple S-subalgebras 
of the semisimple regular subalgebras of $\mathfrak e_6$, treating each 
semisimple ideal independently. The semisimple S-subalgebras are described 
for the classical 
algebras up to rank 6 in \cite{Lorente}, which is enough to construct all 
the semisimple R-subalgebras of $\mathfrak e_6$. However, we only need to 
treat the R-subalgebras $\mathfrak h$ where the regular 
subalgebras $\mathcal R(\mathfrak h)$ lead to an anomaly problem, because
the other cases are protected against anomalies. 
The computation is given in Table \ref{Computation_e6_ssimple}, using 
the fact that one ideal leads to an anomaly or computing the elements 
of the center as described before. Note that for the nonsimple S-subalgebra 
$A_2 \oplus A_1 \subset A_5$, $A_2$ is actually embedded in 
$A_2 \oplus A_2$ \cite{Lorente}, so the question of the two inequivalent 
embeddings $\iota_1$ and $\iota_2$ arises also here, as in \ref{SimpleR}. 
Working by decreasing rank, we have excluded some algebras from this Table 
since they are protected by the ones of higher rank that do not have anomalies.

\eject

\noindent Putting all that together, we obtain the following result:

\begin{prop}
The untwisted coset models with $\mathfrak{g} = \mathfrak{e}_6$ and any
nonregular nonsimple semisimple subalgebra 
$\mathfrak{h}$ do not have global  gauge anomaly
for $k \in \mathbb{Z}$, except for the S-subalgebra $\mathfrak{h} =
\mathfrak{g}_2 \oplus A_2$ and the R-subalgebras appearing in 
Table \ref{Computation_e6_ssimple} with the condition $k\in 3\mathbb Z$
which exhibit global gauge anomaly for $k\in\mathbb Z\setminus3\mathbb Z$.
\end{prop}

\subsection{Examples of nonregular subalgebras of classical Lie algebras}

The semisimple nonregular subalgebra of classical algebra have been 
classified explicitly in \cite{Lorente} only up to rank 6. The general 
classification proposed by Dynkin in \cite{Dynk2} is less explicit and 
does not allow us to treat the anomaly problem in a general form as for 
regular subalgebras. Here we only give some example of classical algebras, 
but the method is always the same once the explicit embedding of a 
subalgebra is known : as for $\mathfrak{e}_6$, we need to look how 
the embedding of the generating element of the center of the considered 
subalgebra is compatible with the coroot lattice of the ambient algebra.

\paragraph{Nonregular semisimple subalgebras of $A_4$.}

The coroot lattice of $A_4$ is given by
\begin{equation}
  P^\vee(A_4) = \left\lbrace \left. \left( \dfrac{a}{5} + q_1, \dots , 
\dfrac{a}{5} + q_4, -\dfrac{4a}{5} - q_1 - \dots - q_4\right) \right| a, 
q_1, \dots q_4 \in \mathbb Z \right\rbrace
\end{equation}
and the coweight lattice $Q^\vee(A_4)$ is given by the same formula but 
with $a = 0$. According to \cite{Lorente}, $A_4$ admits two S-subalgebras 
which are simple : $A_1$ and $B_2$. For $\mathfrak h = A_1$, the embedding 
of the generating element $\lambda^\vee$ of the center of the corresponding 
group is given by
\begin{equation}
  \iota(\lambda^\vee) = (2,1,0,-1,-2) \in Q^\vee(A_4)
\end{equation}
so the quantity $k$ tr$(M\tilde M)$ will be integral for every $k \in 
\mathbb Z$ and there will be no anomaly for this model. For $\mathfrak{h} 
= B_2$, one have
\begin{equation}
   \iota(\lambda^\vee) = (1,0,0,0,-1) \in Q^\vee(A_4)
\end{equation}
which leads to the same conclusion. As we have seen in the regular case, 
all regular subalgebras of $A_4$ (except $A_4$) leads to non-anomalous 
models. We immediately conclude that all the R-subalgebra of $A_4$ are 
protected by their regular $\mathcal R(\mathfrak h)$, so there is also 
no anomaly for these models. Finally, the only anomalous models corresponding 
to $\mathfrak g = A_4$ and an arbitrary semisimple subalgebra are those with 
$\mathfrak h = \mathfrak g$, $Z = \tilde Z \cong Z_5$ and $k \in \mathbb Z 
\setminus 5\mathbb Z$. 

\paragraph{S-subalgebras of $A_5$.}

The coroot lattice of $A_5$ is given by
\begin{equation}
  P^\vee(A_5) = \left\lbrace \left. \left( \dfrac{a}{6} + q_1, \dots , 
\dfrac{a}{6} + q_5, -\dfrac{5a}{6} - q_1 - \dots - q_5\right) \right| a, 
q_1, \dots q_5 \in \mathbb Z \right\rbrace
\end{equation}
and the coweight lattice $Q^\vee(A_5)$ is given by the same formula 
but with $a = 0$. According to \cite{Lorente}, $A_5$ admits six 
S-subalgebras : $A_1$, $A_2$, $A_3$, $C_3$, $A_1 \oplus A_1$ and 
$A_2 \oplus A_1$. For $\mathfrak{h}=A_1$, one has   
\begin{equation}
\iota(\lambda^\vee) = \left( \dfrac{5}{2},\dfrac{3}{2},\dfrac{1}{2},
-\dfrac{1}{2},-\dfrac{3}{2},-\dfrac{5}{2} \right),
\end{equation}
see Table VI of \cite{Lorente}, whereas for $\mathfrak h = A_2,\,A_3$ 
and $C_3$, one has 
\begin{eqnarray}
\iota(\lambda^\vee) = \left( \dfrac{2}{3},\dfrac{2}{3},-\dfrac{1}{3},
\dfrac{2}{3},-\dfrac{1}{3},-\dfrac{1}{3} \right),\ 
\left( \dfrac{1}{2},\dfrac{1}{2},-\dfrac{1}{2},
\dfrac{1}{2},-\dfrac{1}{2},-\dfrac{1}{2} \right),\ 
\left( \dfrac{1}{2},\dfrac{1}{2},\dfrac{1}{2},
-\dfrac{1}{2},-\dfrac{1}{2},-\dfrac{1}{2} \right),
\end{eqnarray}
respectively, see the last 3 entries of Table 5 above.
In all 4 cases, $\iota(\lambda^\vee)\in P^\vee(A_5) \setminus Q^\vee(A_5)$.
Taking $\iota(\lambda^\vee) = \tilde M$ with $\tilde a = 3,4,3,3$, respectively,
and appropriate $\tilde q_i$, and $M\in P^\vee(A_5)$ such that 
$e^{2\ii\pi M}\in Z\cong\mathbb{Z}_p$, we obtain
  \begin{equation}
    \text{tr}(M\tilde M) = \dfrac{5 a \tilde{a}}{p} + n\,,
  \end{equation}
where $n \in \mathbb Z$. There will be no anomaly for $k$ such that 
$k\,{\rm tr}(M\tilde M) \in \mathbb Z$. 
For $\tilde a=3$, this imposes on $k$ the same 
restrictions that the admissibility conditions (\ref{Consistency_Ar}), 
so that the untwisted coset theories corresponding to the S-subalgebras 
$\mathfrak{h}=A_1,A_3,C_3\subset A_5$ do not have anomalies.
\,For the S-subalgebra $\mathfrak{h}=A_2$, we obtain
the non-anomalous models with admissible levels for 
\begin{equation}
  k \in \left\lbrace 
    \begin{array}{ll}
\mathbb Z \cap 2 \mathbb Z = 2 \mathbb Z &\text{if } Z\cong\mathbb Z_2\\      
3 \mathbb Z   &\text{if } Z\cong\mathbb Z_3\\
3 \mathbb Z \cap 2 \mathbb Z = 6 \mathbb Z &\text{if } Z\cong\mathbb Z_6      
    \end{array}\right.
\end{equation}
The other untwisted models corresponding to the S-subalgebra 
$\mathfrak h = A_2\subset A_5$ and non-trivial subgroups $Z$ are anomalous.
\vskip 0.1cm

There are no conceptual or technical difficulties to obtain the no-anomaly 
conditions on $k$ for other subalgebras of $A_5$, and also for other 
classical algebra $\mathfrak g$, once the embeddings are known, but there 
is no general result so each case has to be treated separately. The previous 
examples show that different anomaly conditions could appear according to 
the subalgebra considered.

\section{Conclusions}

We have studied above the conditions for the absence of global gauge
anomaly in the coset models of conformal field theory derived from
WZW models with connected simple compact groups $G=\tilde G/Z$ as the targets  
by gauging a subgroup of the rigid adjoint or twisted-adjoint symmetries
$\,G\ni g\mapsto hg\hspace{0.03cm}\omega(h)^{-1}\in G$, \,where $\omega$ is 
a, possible trivial, automorphism of $G$. \,The full group of such symmetries 
is equal to $\tilde G/Z^\omega$, where $Z^\omega$ is the maximal subgroup 
of the center $\tilde Z$ of the universal covering group $\tilde G$ 
of $G$ for which the (twisted) adjoint action 
is well defined. We considered both the coset models where the full group 
$\tilde G/Z^\omega$ was gauged and the ones where the gauging concerned only 
a closed connected subgroup of $\tilde G/Z^\omega$. Global gauge 
anomalies obstructing the invariance of the Feynman amplitudes of the theory 
under ``large'' gauge transformations non-homotopic to unity may appear 
only for non-simply connected groups $G$ corresponding to Lie algebras 
$\mathfrak{g}$ of types $A_r,\,D_r$ and $\mathfrak{e}_6\,$ (that are all
simply-laced). Using the results
\cite{Dynkin,Lorente,Minchenko} on the classification of semisimple 
Lie subalgebras of simple Lie algebras, we obtained a complete 
list of non-anomalous coset models (without boundaries) for groups $G$ 
with the Lie algebra $A_r,\,D_r$ or $\mathfrak{e}_6$ if the gauged symmetry 
subgroup $\subset\tilde G/Z^\omega$ corresponds to a regular Lie subalgebra 
$\mathfrak{h}\subset\mathfrak{g}$ or, for $\mathfrak{g}=\mathfrak{e}_6$,
to any semisimple Lie subalgebra. The global gauge anomalies that appear
in the other coset model should render them inconsistent on the quantum
level, as was argued in \cite{GSW}. 
\eject

\noindent{\bf\Large Appendices}
\vskip -0.5cm

\appendix

\section{Gauge-invariance condition}\label{app:1}

Here we prove the equivalence between relations (\ref{gaugeinv}) and
(\ref{eval}). From Eq.\,(\ref{gauged_action}), we have to show that
\begin{eqnarray}
&&\frac{_k}{^{4\pi}}\int{\rm tr}\,\big(({}^h\hspace{-0.05cm}g^{-1}d
\hspace{0.02cm}
{}^h\hspace{-0.05cm}g)\hspace{0.04cm}\omega({}^h\hspace{-0.07cm}A)
+(d\hspace{0.02cm}{}^h\hspace{-0.05cm}g){}^h\hspace{-0.05cm}g^{-1}
\hspace{-0.05cm}\hspace{0.04cm}{}^h\hspace{-0.08cm}A+{}^h
\hspace{-0.05cm}g^{-1}\hspace{0.04cm}
{}^h\hspace{-0.08cm}A\hspace{0.03cm}{}^h\hspace{-0.05cm}g\hspace{0.05cm}
\omega({}^h\hspace{-0.05cm}A)\big)\cr
&&-\,\frac{_k}{^{4\pi}}\int{\rm tr}\,\big((g^{-1}dg)\hspace{0.02cm}\omega(A)
+(dg)g^{-1}\hspace{-0.05cm}A+g^{-1}\hspace{-0.05cm}
Ag\hspace{0.05cm}\omega(A)\big)\cr
&=&-\frac{_{k}}{^{4\pi}}\int_\Sigma
{\rm tr}\,\big(g^{-1}dg\hspace{0.05cm}\omega(h^{-1}dh)+(dg)g^{-1}h^{-1}dh
+g^{-1}(h^{-1}dh)g\hspace{0.05cm}\omega(h^{-1}dh)\big)
\label{toest}
\end{eqnarray}
But
\begin{eqnarray}
&&\frac{_k}{^{4\pi}}\int{\rm tr}\,\big(({}^h\hspace{-0.05cm}g^{-1}d
\hspace{0.02cm}
{}^h\hspace{-0.05cm}g)\hspace{0.04cm}\omega({}^h\hspace{-0.07cm}A)
+(d\hspace{0.02cm}{}^h\hspace{-0.05cm}g){}^h\hspace{-0.05cm}g^{-1}
\hspace{-0.05cm}\hspace{0.04cm}{}^h\hspace{-0.08cm}A+{}^h\hspace{-0.05cm}
g^{-1}\hspace{0.04cm}
{}^h\hspace{-0.08cm}A\hspace{0.03cm}{}^h\hspace{-0.05cm}g\hspace{0.05cm}
\omega({}^h\hspace{-0.05cm}A)\big)\cr
&&-\,\frac{_k}{^{4\pi}}\int{\rm tr}\,\big((g^{-1}dg)\hspace{0.02cm}\omega(A)
+(dg)g^{-1}\hspace{-0.05cm}A+g^{-1}\hspace{-0.05cm}
Ag\hspace{0.05cm}\omega(A)\big)\cr
&=&\frac{_k}{^{4\pi}}\int{\rm tr}\,
\big(\omega(h)g^{-1} h^{-1}\,d(h\hspace{0.02cm}g
\hspace{0.04cm}\omega(h)^{-1})\hspace{0.04cm}\omega(h\hspace{-0.01cm}
A\hspace{0.02cm}h^{-1}+hdh^{-1})\cr
&&\qquad\qquad+d(hg\omega(h)^{-1})\hspace{0.04cm}\omega(h)\hspace{0.02cm}g^{-1}
\hspace{0.02cm}h^{-1}\hspace{0.02cm}(hAh^{-1}+hdh^{-1})\cr
&&\qquad\qquad+\,\omega(h)\hspace{0.02cm}
g^{-1} h^{-1}(hAh^{-1}+hdh^{-1})\hspace{0.03cm}h\hspace{0.04cm}g\hspace{0.04cm}
\omega(h)^{-1}\hspace{0.05cm}
\omega(hAh^{-1}+hdh^{-1})\big)\cr
&&-\,\frac{_k}{^{4\pi}}\int{\rm tr}\,\big((g^{-1}dg)\hspace{0.02cm}\omega(A)
+(dg)g^{-1}\hspace{-0.05cm}A+g^{-1}\hspace{-0.05cm}
Ag\hspace{0.05cm}\omega(A)\big)\cr
&=&\frac{_k}{^{4\pi}}\int{\rm tr}\,
\big(g^{-1}\hspace{0.02cm}((h^{-1}dh)\hspace{0.02cm}g
\hspace{0.04cm}+(dg)\hspace{0.02cm}
-g\hspace{0.04cm}\omega(h^{-1}dh))\hspace{0.05cm}\omega(
A+(dh^{-1})h)\cr
&&\qquad\qquad+((dh)g\hspace{0.02cm}+h(dg)
\hspace{0.02cm}-hg\hspace{0.02cm}\omega(h^{-1}dh))
\hspace{0.06cm}g^{-1}
\hspace{0.02cm}h^{-1}\hspace{0.02cm}(hAh^{-1}+hdh^{-1})\cr
&&\qquad\qquad+\,\omega(h)\hspace{0.02cm}
g^{-1} h^{-1}(hAh^{-1}+hdh^{-1})\hspace{0.03cm}hg\hspace{0.04cm}\omega(h)^{-1}
\hspace{0.05cm}
\omega(hAh^{-1}+hdh^{-1})\big)\cr
&&-\,\frac{_k}{^{4\pi}}\int{\rm tr}\,\big((g^{-1}dg)\hspace{0.02cm}\omega(A)
+(dg)g^{-1}\hspace{-0.05cm}A+g^{-1}\hspace{-0.05cm}
Ag\hspace{0.05cm}\omega(A)\big)\cr
&=&\frac{_k}{^{4\pi}}\int{\rm tr}\,
\big(g^{-1}\hspace{0.02cm}(h^{-1}dh)\hspace{0.02cm}g
\hspace{0.04cm}\omega(
A-h^{-1}dh)-(g^{-1}dg)\hspace{0.04cm}\omega(h^{-1}dh)
-\hspace{0.04cm}\omega(h^{-1}dh)\hspace{0.05cm}\omega(
A)\cr
&&\qquad\qquad+(h^{-1}dh)\hspace{0.02cm}(A-h^{-1}dh)-(dg)
\hspace{0.02cm}g^{-1}
\hspace{0.02cm}(h^{-1}dh)-g\hspace{0.04cm}\omega(h^{-1}dh)g^{-1}
\hspace{0.02cm}(A-h^{-1}dh)
\hspace{0.06cm}\cr
&&\qquad\qquad-\,\hspace{0.02cm}
g^{-1}(h^{-1}dh)\hspace{0.04cm}g
\hspace{0.05cm}
\omega(A-h^{-1}dh)
-g^{-1}(A-h^{-1}dh)\hspace{0.04cm}g
\hspace{0.05cm}
\omega(h^{-1}dh)-g^{-1}(h^{-1}dh)g\hspace{0.04cm}\omega(h^{-1}dh)\big)\cr
&=&\frac{_k}{^{4\pi}}\int{\rm tr}\,
\big(-(g^{-1}dg)\hspace{0.04cm}\omega(h^{-1}dh)
-(dg)\hspace{0.02cm}g^{-1}
\hspace{0.02cm}(h^{-1}dh)-g^{-1}(h^{-1}dh)g\hspace{0.04cm}\omega(h^{-1}dh)\big)
\end{eqnarray}
which establishes identity (\ref{toest}).

\section{Arithmetical properties}\label{app:2}

For $a,b \in \mathbb{Z}$ we denote $a \wedge b$ the greatest common divisor and
$a \vee b$ the least common multiple of $a$ and $b$.

\begin{prop}\label{lcm}
 Let $k_1, \ldots k_s \in \mathbb{Z}$ and $k \in \mathbb{Z}$ such that $\forall
i = 1 \ldots s, k \in k_i \mathbb{Z}$, then
\begin{equation}
 k \in \left( k_1 \vee \cdots \vee k_s \right) \mathbb{Z}
\end{equation}
\end{prop}
The demonstration is done by induction on $s$.

\begin{prop}\label{lcmfrak}
Let $k_1, \ldots k_s \in \mathbb{Z}$ such that $\forall i = 1 \ldots s$, $k_i =
\dfrac{a}{a \wedge b_i}$ with $a, b_1, \ldots, b_s \in \mathbb{Z}$, then
\begin{equation}
k_1 \vee \ldots \vee k_s = \dfrac{a}{a \wedge b_1 \wedge \cdots \wedge b_s}
\end{equation}
\end{prop}

The demonstration is done by induction on $s$:
\begin{itemize}
 \item $s=2$ 
\begin{equation}
 \dfrac{a}{a\wedge b_1} \vee \dfrac{a}{a\wedge b_2} =
\dfrac{\dfrac{a^2}{(a\wedge b_1)(a\wedge b_2)}}{\dfrac{a}{a\wedge b_1} \wedge
\dfrac{a}{a\wedge b_2}}
\end{equation}
using $ab = (a \wedge b)(a \vee b)$. Then we can rewrite the denominator:
\begin{equation}
 \dfrac{a}{a\wedge b_1} \wedge \dfrac{a}{a\wedge b_2} = a \dfrac{(a \wedge b_2)
\wedge (a\wedge b_1)}{(a \wedge b_2) (a\wedge b_1)},
\end{equation}
thus
\begin{equation}
\dfrac{a}{a\wedge b_1} \vee \dfrac{a}{a\wedge b_2} = \dfrac{a}{a\wedge b_1
\wedge a \wedge b_2} =\dfrac{a}{a\wedge b_1 \wedge b_2}.
\end{equation}

\item Suppose the result true for $s \geq 2$, the result for $s+1$ is trivially
true, using the induction hypothesis at rank $s$, then $2$.

\end{itemize}

\end{document}